\documentclass[12pt]{article}

\usepackage{yurie}
\usepackage{yurie-math}
\usepackage{yurie-phy}

\numberwithin{equation}{section}

\allowdisplaybreaks

\NewDocumentCommand{\includeImage}{m g}{
    \begin{figure}[htbp]
        \centering
        \includegraphics[trim=1.5cm 1.5cm 1.5cm 1.5cm,clip]{./img/pdf/#1.pdf}
        \IfNoValueTF{#2}{\caption{#1}}{\caption{#2}}
        \label{fig: #1}
    \end{figure}
}

\newcommand{\reg}{\texttt{R}}
\newcommand{\mlimit}{N}

\newcommand{\DeltaS}{\wave{\Delta}}

\newcommand{\halfdim}{\frac{d}{2}}

\newcommand{\coTwo}{c}
\newcommand{\coThree}{C}
\newcommand{\coOPE}{C}
\newcommand{\coShadow}{S}

\newcommand{\scalar}{\phiup}
\newcommand{\gluon}{\mathsf{g}}
\newcommand{\graviton}{\mathsf{h}}

\newcommand{\TwoToN}{\ensuremath{2 \to n}\xspace}
\newcommand{\OneToN}{\ensuremath{1 \to n}\xspace}
\newcommand{\OneToThree}{\ensuremath{1 \to 3}\xspace}
\newcommand{\TwoToOne}{\ensuremath{2 \to 1}\xspace}
\newcommand{\TwoToTwo}{\ensuremath{2 \to 2}\xspace}
\newcommand{\TwoToThree}{\ensuremath{2 \to 3}\xspace}

\begin{document}

\begin{titlepage}

    \title{Shadow Completion in Celestial OPEs}

    \author{Reiko Liu$^{b}$, Zijian Liu$^{c}$, Wen-Jie Ma$^{a,b}$}

    \date{}

    \maketitle\thispagestyle{empty}

    \affiliation{a}{Fudan Center for Mathematics and Interdisciplinary Study, Fudan University, Shanghai, 200438, China}

    \affiliation{b}{Shanghai Institute for Mathematics and Interdisciplinary Sciences (SIMIS), Shanghai, 200433, China}

    \affiliation{c}{Department of Physics and Center for Field Theory and Particle Physics, Fudan University, Shanghai 200438, China}

    \email{
        reiko.antoneva@foxmail.com,
        liuzj24@m.fudan.edu.cn,
        wenjie.ma@simis.cn
    }

    \vfill

    \begin{abstract}
        We argue that celestial OPEs must be supplemented by shadow-basis operators. Although the shadow transform does not introduce new bulk degrees of freedom, it provides a distinct primary state in the boundary celestial theory.
        From OPE consistency, we show that the ordinary celestial OPE does not close on Mellin-basis exchanges alone. Rather, the same exchanged bulk particle must also appear through its shadow-basis representative. This leads to a shadow-completed OPE, with the shadow OPE coefficient fixed by the ordinary collinear coefficient through the universal shadow factor.
        We discuss the corresponding boundary Hilbert-space interpretation, extend this argument to gluons and gravitons, and verify the shadow exchange directly in tree-level regular celestial amplitudes, including a scalar $\TwoToN$ analysis and an explicit five-point example.
    \end{abstract}

    \vfill

\end{titlepage}

%%%%%%%%%%%%%%%%%%%%%%%%%%%%%%%%%%%%%%%%%%%%%%%%%%%%%%%%%%%%%%%%
%%%%%%%%%%%%%%%%%%%%%%%%%%%%%%%%%%%%%%%%%%%%%%%%%%%%%%%%%%%%%%%%

\begingroup
\hypersetup{linkcolor=black}
\tableofcontents
\endgroup

%%%%%%%%%%%%%%%%%%%%%%%%%%%%%%%%%%%%%%%%%%%%%%%%%%%%%%%%%%%%%%%%

\clearpage

%%%%%%%%%%%%%%%%%%%%%%%%%%%%%%%%%%%%%%%%%%%%%%%%%%%%%%%%%%%%%%%%

\newpage

%%%%%%%%%%%%%%%%%%%%%%%%%%%%%%%%%%%%%%%%%%%%%%%%%%%%%%%
\section{Introduction}
%%%%%%%%%%%%%%%%%%%%%%%%%%%%%%%%%%%%%%%%%%%%%%%%%%%%%%%

Celestial holography reformulates four-dimensional scattering amplitudes as correlation functions on the celestial sphere \cite{Cheung:2016iub,Pasterski:2016qvg,Pasterski:2017kqt,Strominger:2017zoo,Raclariu:2021zjz,Pasterski:2021rjz,Pasterski:2021raf,McLoughlin:2022ljp}. For massless external particles, the standard construction employs the Mellin conformal basis: the external energy is Mellin-transformed, yielding a two-dimensional conformal primary operator $\mathcal{O}_{\Delta,J}(z,\bar z)$. This basis diagonalizes four-dimensional boosts and realizes the Lorentz group as the global conformal group on the celestial sphere.

A given bulk massless particle, however, admits another natural conformal primary representative. The shadow transform maps a primary of conformal dimension $\Delta$ and spin $J$ to another primary with quantum numbers
\begin{equation}
    (\Delta,J)\longmapsto (2-\Delta,-J)
    \, .
\end{equation}
Thus a single four-dimensional massless particle has two natural celestial representatives: the Mellin-basis operator and its shadow-basis counterpart \cite{Crawley:2021ivb,Fan:2021isc,Fan:2021pbp,Chang:2022jut,Chang:2022seh,Furugori:2023hgv,Surubaru:2025qhs,Himwich:2025bza,Pranzetti:2025flv,Liu:2025dhh,Sharma:2021gcz,Jorge-Diaz:2022dmy,Banerjee:2022hgc,Banerjee:2024hvb,Banerjee:2025oyu,Navarro:2026rna}. This raises a basic question: when mapping a bulk external particle to the boundary, should it be represented by the Mellin basis, by the shadow basis, or by both? Equivalently, what is the correct boundary state space associated with a single bulk one-particle state?

This question is intimately tied to the spectrum of the celestial OPE.\footnote{The celestial OPE has been extensively studied in \eg \cite{Fan:2019emx,Pate:2019lpp,Lam:2017ofc,Garcia-Sepulveda:2022lga,Atanasov:2021cje,Chang:2021wvv,Fan:2022kpp,Chang:2023ttm,Fan:2023lky,Liu:2024lbs,Liu:2024vmx,Fan:2021isc,Fan:2021pbp,Chang:2022jut,Chang:2022seh,Surubaru:2025qhs,Himwich:2025bza,Pranzetti:2025flv,Liu:2025dhh,Sleight:2023ojm,Iacobacci:2024nhw,Malherbe:2025qex,Liu:2025voe}.} The ordinary collinear OPE of two massless Mellin-basis operators is typically derived from the collinear limit of scattering amplitudes. For instance, in massless $\phi^3$ theory one finds
\begin{equation}
    \scalar^{\inn}_{\Delta_1}
    \scalar^{\inn}_{\Delta_2}
    \sim
    % \frac{1}{4}\bt*{\Delta_1-1,\Delta_2-1}
    \scalar^{\inn}_{\Delta_{12}-2}
    % +\cdots
    \, .
\end{equation}
At first sight, this suggests that the OPE closes on Mellin-basis operators alone. The central point of this paper is that this conclusion is incomplete. We argue that even if one begins with an operator algebra generated solely by Mellin-basis primaries, consistency of the ordinary OPE naturally forces the appearance of shadow-basis operators.

The mechanism is most transparent from OPE consistency. Consider a celestial three-point function describing the decay of a massive scalar into two massless scalars $\vev{\scalar^{\inn,m}_{\Delta_1}\scalar^{\out}_{\Delta_2}\scalar^{\out}_{\Delta_3}}$.
This correlator takes the standard form of a two-dimensional scalar three-point function and is non-vanishing when the two outgoing celestial points are separated. Taking $12$ OPE limit should reduce it to a two-point function between the exchanged operator and the remaining massless operator. If the exchanged operator were solely an ordinary Mellin-basis massless scalar, this two-point function would be contact-supported on the celestial sphere and would vanish at separated points. Hence the ordinary Mellin exchange cannot reproduce the non-contact behavior of the original three-point function.
The missing bridge is supplied by the mixed two-point function between a Mellin-basis operator and its shadow, forcing the OPE to contain the shadow-basis representative of the same bulk massless particle.

This leads us to propose the shadow-completed celestial OPE. In the scalar case, the ordinary Mellin exchange must be supplemented by its shadow counterpart:
\begin{equation}
    \scalar^{\inn}_{\Delta_1}
    \scalar^{\inn}_{\Delta_2}
    \sim
    % \mathcal{C}^{\rm M}_{\Delta_{12}-2}
    \scalar^{\inn}_{\Delta_{12}-2}
    +
    % \mathcal{C}^{\rm Sh}_{\Delta_{12}-2}
    \shadow{\scalar^{\inn}_{\Delta_{12}-2}}
    % +\cdots
    \, .
\end{equation}
The shadow operator in this expression does not represent a new bulk particle. It an alternative celestial conformal-primary representative of the same four-dimensional massless scalar state. The crucial point is that the local boundary OPE does not close on the Mellin representative alone. The same logic applies to purely massless correlators after introducing a mass regularization, or equivalently by working with regular celestial amplitudes \cite{Chang:2022seh,Liu:2025dhh,Liu:2025voe,Liu:2026ocv}, and extends to higher-point scalar correlators by successive reduction through ordinary collinear OPEs to effective three-point functions.

This observation also clarifies the boundary Hilbert space interpretation. Although the shadow basis is related to the Mellin basis by a nonlocal integral transform and thus does not represent a new copy of the bulk one-particle Hilbert space, the local state
\begin{equation}
    \shadow{\scalar_{\Delta}}(0)\vac
\end{equation}
is not contained in the ordinary Verma module generated by $\scalar_{\Delta}(0)\vac$ and its descendants. In other words, the shadow state cannot be represented as a convergent linear combination of the states in this Verma module. Consequently, while the bulk particle content remains unchanged, the local boundary operator algebra must be enlarged from a purely Mellin-basis module to a shadow-completed module.

We further determine the shadow exchange coefficient. This coefficient is not an independent OPE datum but is fixed by the ordinary collinear OPE coefficient and by the universal shadow factor relating a three-point structure to the shadow transform of one of its legs.

The same reasoning applies to massless particles with spin. For gluons and gravitons, the ordinary collinear OPE produces a Mellin-basis exchanged operator, but its ordinary Mellin two-point function remains contact-supported. Consequently, the separated OPE limit again requires the shadow-basis representative of the exchanged gluon or graviton.

As an independent check, we derive the same shadow exchange directly from tree-level regular celestial amplitudes. We analyze $\TwoToN$ massless scalar scattering within the regular celestial amplitude framework. Using a single-particle mass regularization and the split representation developed in \cite{Chang:2023ttm,Liu:2024vmx,Liu:2024lbs}, we extract the OPE behavior in the $12$ channel. We find that there exists an operator of dimension $4-\Delta_{12}$, precisely the shadow dimension of the ordinary collinear exchange $\Delta_{12}-2$. The coefficient extracted from this direct $n$-point amplitude analysis agrees with the shadow OPE coefficient predicted by the OPE-consistency argument.

The remainder of this paper is organized as follows. In Section \ref{sec:conventions}, we summarize our conventions for celestial kinematics, conformal structures, and shadow transforms. In Section \ref{sec:shadow_states}, we present the OPE-consistency argument for shadow exchange, beginning with massive decay and extending to purely massless and higher-point scalar correlators. We also discuss the status of the corresponding shadow states in the boundary Hilbert space, derive the relation between shadow and Mellin OPE coefficients, and extend the discussion to gluon and graviton OPEs. In Section \ref{sec:tree_level_exchange}, we rederive the shadow exchange directly from $n$-point tree-level regular celestial amplitudes. Finally, in Section \ref{sec:five_point_check}, we explicitly compute a five-point regular celestial amplitude to verify the shadow exchange predicted by the OPE analysis.

%%%%%%%%%%%%%%%%%%%%%%%%%%%%%%%%%%%%%%%%%%%%%%%%%%%%%%%%%%%%%%%%

\section{Conventions}
\label{sec:conventions}

We use the conventions of \cite{Liu:2025voe}; the notation needed below is summarized here.

\paragraph{Celestial kinematics.}
\begin{itemize}
    \item
        The bulk spacetime is four-dimensional Lorentzian with metric signature $(-,+,+,+)$, and the celestial sphere is two-dimensional Euclidean.
    \item
        The label $\inn$ and $\out$ denote incoming and outgoing states or operators, respectively.
    \item
        Boundary coordinates are $(z,\zb)$, and the conformal weights $(h,\hb)$ are related to the conformal dimension and spin by $h=(\Delta+J)/2$ and $\hb=(\Delta-J)/2$. Unless stated otherwise, we write only the holomorphic part and leave the antiholomorphic part implicit.
        For sums and differences of these quantities, we use the multi-label notation in the style of
        \begin{equation}
            \Delta_{a_{1}\cdots a_{n}}
            \equiv
            \sum_{i=1}^{n}\Delta_{a_{i}}
            \, ,
            \quad
            \Delta_{a_{1}\cdots a_{n},b_{1}\cdots b_{m}}
            \equiv
            \sum_{i=1}^{n}\Delta_{a_{i}}-
            \sum_{j=1}^{m}\Delta_{b_{j}}
            \, .
        \end{equation}
    \item
        We denote massless, massive and tachyonic momenta by $q$, $p$ with $p^2=-m^2$, and $k$ with $k^2=m^2$. The parametrizations are
        \begin{align}
            \label{eq: momentum parametrization}
            &
            q=\omega\hat{q}
            =
            \omega(1+z \zb,z+\zb,-i (z-\zb),1-z \zb)
            \, ,
            \textInMath{for} \omega\geq 0
            \, ,
            \\
            &
            p=m\phat
            =
            \frac{m}{2y}(1+z \zb+y^2,z+\zb,-i (z-\zb),1-z \zb-y^2)
            \, ,
            \textInMath{for} y>0
            \, ,
            \\
            &
            k=m\khat
            =
            \frac{m}{2y}(1+z \zb-y^2,z+\zb,-i (z-\zb),1-z \zb+y^2)
            \, ,
            \textInMath{for} y\in\RR
            \, .
        \end{align}
\end{itemize}

\paragraph{Special functions.}
We use the following shorthand for Beta and Gamma functions:
\begin{align}
    \label{eq: special function notation}
    \bt*{a,b}
    &\equiv
    \frac{\gm{a}\gm{b}}{\gm{a+b}}
    \, ,
    \\
    \gm{a_{1},\dotsc,a_{n}}
    &\equiv
    \prod_{i=1}^{n}\gm{a_{i}}
    \, ,
    \\
    \mg*{a_{1},\dotsc,a_{n}}{b_{1},\dotsc,b_{m}}
    &\equiv
    \left.
    \prod_{i=1}^{n}\gm{a_{i}}
    \middle/
    \prod_{j=1}^{m}\gm{b_{j}}
    \right.
    \, .
\end{align}

\paragraph{CFT kinematics.}
\begin{itemize}
    \item
        We write $\op_{i}\eqq \op_{\Delta_{i},J_{i}}(z_{i})$ when no confusion can arise. Double brackets denote the kinematic conformal structures:
        \begin{equation}
            \label{eq: conformal structure}
            \begin{aligned}
            &
            \vevv{\op_{1}\op_{2}}
            =
            % \delta_{\Delta_{1},\Delta_{2}}\delta_{J_{1},J_{2}}
            z_{1,2}^{-2 h_1}
            \zb_{1,2}^{-2 \bar{h}_1}
            \, ,
            \\
            &
            \vevv{\op_{1}\op_{2}\op_{3}}
            =
            z_{1,2}^{h_{3,12}} z_{2,3}^{h_{1,23}} z_{1,3}^{h_{2,13}}
            \zb_{1,2}^{\bar{h}_{3,12}} \zb_{2,3}^{\bar{h}_{1,23}} \zb_{1,3}^{\bar{h}_{2,13}}
            \, .
            \end{aligned}
        \end{equation}
        Here $z_{i,j}\equiv z_i-z_j$. The two- and three-point coefficients are defined by
        \begin{equation}
            \vev{\op_{1}\op_{2}}
            =
            \vevv{\op_{1}\op_{2}}
            \,
            \coTwo(\op_{1}\op_{2})
            \, ,
            \quad
            \vev{\op_{1}\op_{2}\op_{3}}
            =
            \vevv{\op_{1}\op_{2}\op_{3}}
            \,
            \coThree(\op_{1}\op_{2}\op_{3})
            \, .
        \end{equation}
    \item
        The shadow transform acts on a primary operator as
        \begin{equation}
            \label{eq: shadow transform}
            \wave{\op}_{\wave{\Delta},\wave{J}}(z)
            \eqq
            \shadow[\op_{\Delta,J}] (z)
            \eqq
            \intt{d^{2}z'}
            (z-z')^{2h-2}
            (\zb-\zb')^{2\hb-2}
            \op_{\Delta,J}(z',\zb')
            \, .
        \end{equation}
        The shadow operator has conformal dimension $\wave{\Delta}\eqq 2-\Delta$ and spin $\wave{J}\eqq -J$.
        For generic $\Delta\in\CC$, the inverse transform reads
        \begin{equation}
            \label{eq:inverse_shadow_transform}
            \op_{\Delta,J} (z)
            =
            N_{\Delta,J}
            \intt{d^{2}z'}
            (z-z')^{-2h}
            (\zb-\zb')^{-2\hb}
            \shadow[\op_{\Delta,J}](z')
            \, ,
        \end{equation}
        where $N_{\Delta,J} = -\pi^{-2}(\Delta -J-1) (\Delta +J-1)$.
    \item
        The star-triangle relation is
        \begin{equation}
            \label{eq: star-triangle relation}
            \intt{d^{2}z'}
            (z-z')^{2h-2}
            (\zb-\zb')^{2\hb-2}\vevv{\op_{1}\op_{2}\op_{\Delta_{3},J_{3}}(z')}
            =
            \coShadow_{\Delta_3,J_3}^{\Delta_1,J_1,\Delta_2,J_2}
            \vevv{\op_{1}\op_{2}\op_{\wave\Delta_{3},\wave{J}_{3}}(z)}
            \, ,
        \end{equation}
        and the prefactor is called the shadow factor:
        \begin{align}\label{eq:shadow_factor}
            \coShadow_{\Delta_3,J_3}^{\Delta_1,J_1,\Delta_2,J_2}
            =
            \pi
            \mg{
                2 h_3-1,
                h_{2,13}+1,
                \bar{h}_{1,23}+1
            }{
                2-2 \bar{h}_3,
                h_{23,1},
                \bar{h}_{13,2}
            }
            \, .
        \end{align}
\end{itemize}

%%%%%%%%%%%%%%%%%%%%%%%%%%%%%%%%%%%%%%%%%%%%%%%%%%%%%%%%%%%%%%%%%%%%%%%%%
\section{Shadow states in CCFT}\label{sec:shadow_states}
%%%%%%%%%%%%%%%%%%%%%%%%%%%%%%%%%%%%%%%%%%%%%%%%%%%%%%%%%%%%%%%%%%%%%%%%
\subsection{Shadow exchange from OPE consistency}
%%%%%%%%%%%%%%%%%%%%%%%%%%%%%%%%%%%%%%%%%%%%%%%%%%%%%%%%%%%%%%%%%%%%%%%%%
In this section we argue that shadow operators are naturally forced upon us, even if one starts from an operator algebra generated only by Mellin-basis primaries. The basic point can already be seen from the consistency of the ordinary OPE with the bulk two-point pairing in a simple three-point example. Consider the celestial three-point function describing the decay of a massive scalar into two massless scalars,
\begin{align}
    \vev{\scalar^{\inn,m}_{\Delta_1}\scalar^{\out}_{\Delta_2}\scalar^{\out}_{\Delta_3}}
    \, .
\end{align}
This correlator has the standard form of a two-dimensional scalar three-point function and is nonzero for separated celestial points, in particular for $\hat q_2\neq \hat q_3$. We now take the OPE limit $\hat q_1\to \hat q_2$. The leading contribution to the three-point function must reduce to a two-point function of the form
\begin{align}
    \vev{{\cal Y}^{\inn}_{12}\scalar^{\out}_{\Delta_3}}
    \, ,
\end{align}
where $\mathcal{Y}^{\inn}_{12}$ denotes an operator appearing in the $12$-channel OPE which can contribute to the three-point function. A naive possibility is that ${\cal Y}^{\inn}_{12}$ is an ordinary Mellin-basis massless scalar primary $\scalar^{\inn}_{\Delta_0}$. However, the two-point function of ordinary massless scalar celestial primaries is contact-supported on the celestial sphere,
\begin{align}
    \vev{
        \scalar^{\inn}_{\Delta_0}(z_2)
        \scalar^{\out}_{\Delta_{3}}(z_3)
    }
    \propto
    \delta^{(2)}(z_{2,3})
    \, .
\end{align}
This contribution vanishes in the separated region $\hat{q}_2\neq\hat{q}_3$. Thus the ordinary Mellin-basis operator $\scalar^{\inn}_{\Delta_{0}}$ cannot account for the non-contact part of the three-point function. The $12$ OPE must therefore contain a scalar primary ${\cal Y}^{\inn}_{12}$ such that $\vev{\mathcal{Y}^{\inn}_{12}(z_2)\scalar^{\out}_{\Delta_3}(z_3)}$ is nonzero for $\hat{q}_2\neq \hat{q}_3$. By two-dimensional conformal invariance, such a non-contact scalar two-point function must take the standard power-law form.

It remains to identify what this operator ${\cal Y}^{\inn}_{12}$ can be. The non-vanishing of the boundary two-point function should reflect a non-vanishing pairing of the corresponding bulk states. Since $\scalar^{\out}_{\Delta_{3}}$ is obtained by Mellin transforming a four-dimensional outgoing massless scalar state, ${\cal Y}^{\inn}_{12}$ must represent a bulk state which has a nonzero pairing with this outgoing massless scalar particle.
In the bulk theory, a one-particle massless scalar state has a nonzero two-point pairing only with the corresponding one-particle state of the same species and opposite in/out orientation. Therefore ${\cal Y}^{\inn}_{12}$ should not be interpreted as a new bulk degree of freedom. Rather, it should be another celestial representative of the incoming one-particle state of the same massless scalar.

This observation strongly constrains the possible candidate of ${\cal Y}^{\inn}_{12}$. The ordinary Mellin transform gives one conformal primary basis for the massless scalar representation. A second conformal primary basis is obtained by applying the two-dimensional shadow transform to the Mellin primary and the mixed two-point function between a Mellin-basis scalar and its shadow has the standard CFT form,
\begin{align}
    \vev{
        \scalar^{\inn}_{\Delta}(z_2)
        \shadow{\scalar^{\out}_{2-\Delta}}(z_3)
    }=\vev{
        \scalar^{\out}_{\Delta}(z_3)
        \shadow{\scalar^{\inn}_{2-\Delta}}(z_2)
    }
    \propto
    \frac{1}{|z_{2,3}|^{2\Delta}}
    \, .
\end{align}
The natural candidate is therefore
\begin{align}
    {\cal Y}^{\inn}_{12}=\shadow{\scalar^{\inn}_{2-\Delta_{3}}}
    \, .
\end{align}
In other words, the extra operator required by the OPE limit is not associated with a new bulk degree of freedom. It is the shadow-basis representative of the same incoming massless scalar one-particle state.

The above argument can be directly generalized to the purely massless case. Consider the three-point function $\vev{\scalar^{\inn}_{\Delta_1}\scalar^{\inn}_{\Delta_2}\scalar^{\out}_{\Delta_3}}$.
For strictly massless external states, the celestial three-point amplitude is distributional and its singular support obscures the usual OPE interpretation \cite{Pasterski:2017ylz,Chang:2022seh}. In order to use the ordinary CFT OPE language, we introduce a mass regularization, or equivalently work with the regular celestial amplitude introduced in \cite{Liu:2025voe}. In this regularized setup, the massless three-point function takes the standard scalar three-point form and is nonvanishing for separated celestial points. Hence the same OPE consistency argument used above can be applied.

The leading collinear celestial OPE in the $12$ channel contains the massless scalar primary
\begin{align}
    \scalar^{\inn}_{\Delta_1}(z_1)
    \scalar^{\inn}_{\Delta_2}(z_2)
    \sim
    \coOPE_{\Delta_{12}-2}\,
    \scalar^{\inn}_{\Delta_{12}-2}(z_2)
    \, ,
\end{align}
where, in massless $\phi^3$ theory,
\begin{align}\label{eq:scalar_Mellin_OPE_coefficient}
    \coOPE_{\Delta_{12}-2}
    =
    \frac{1}{4}\bt*{\Delta_1-1,\Delta_2-1}
    \, .
\end{align}
If the $12$ OPE contained only the ordinary Mellin-basis operator $\scalar^{\inn}_{\Delta_{12}-2}$, then after taking the OPE limit the remaining two-point function with the third operator would be $\vev{\scalar^{\inn}_{\Delta_{12}-2}(z_2)\scalar^{\out}_{\Delta_3}(z_3)}$.
However, the two-point function of ordinary massless celestial primaries in the Mellin basis is contact-supported on the celestial sphere,
\begin{align}
    \vev{
        \scalar^{\inn}_{\Delta_{12}-2}(z_2)
        \scalar^{\out}_{\Delta_3}(z_3)
    }
    \propto
    \delta^{(2)}(z_{2,3})
    \, .
\end{align}
This contribution vanishes for separated points $\hat q_2\neq \hat q_3$. It therefore cannot reproduce the non-contact power-law behavior of the regular massless three-point function.

Thus, as in the massive-decay example, the $12$ OPE must contain another scalar primary ${\cal X}^{\inn}_{12}$ such that
\begin{align}
    \vev{
        {\cal X}^{\inn}_{12}(z_2)
        \scalar^{\out}_{\Delta_3}(z_3)
    }
\end{align}
is nonzero for $\hat q_2\neq \hat q_3$. By two-dimensional conformal invariance, this non-contact two-point function must take the standard power-law form. Since $\scalar^{\out}_{\Delta_3}$ represents an outgoing massless scalar one-particle state, the operator ${\cal X}^{\inn}_{12}$ should represent the same bulk scalar state with the opposite in/out orientation, but in a different celestial conformal basis. The natural candidate is therefore the shadow-basis representative of the ordinary $12$-channel Mellin operator,
\begin{align}
    {\cal X}^{\inn}_{12}
    =
    \shadow{\scalar^{\inn}_{\Delta_{12}-2}}
    \, .
\end{align}
Indeed, the mixed two-point function between a Mellin-basis scalar and its shadow has the standard CFT form,
\begin{align}\label{eq:mixed_two_point}
    \vev{
        \shadow{\scalar^{\inn}_{\Delta_{12}-2}}(z_2)
        \scalar^{\out}_{4-\Delta_{12}}(z_3)
    }
    \propto
    \frac{1}{|z_{2,3}|^{2(4-\Delta_{12})}}
\end{align}
once the dimensions are matched by the scaling support of the regular three-point amplitude.

The same conclusion also holds inside higher-point scalar correlators. To see this, consider an $n$-point function with two incoming massless scalars and $n-2$ outgoing massless scalars $\vev{\scalar^{\inn}_{\Delta_1}\scalar^{\inn}_{\Delta_2}\scalar^{\out}_{\Delta_3}\cdots\scalar^{\out}_{\Delta_n}}$.
We take a sequence of ordinary collinear OPEs in the outgoing cluster. First, the $(n-1)n$ OPE contains
\begin{align}
    \scalar^{\out}_{\Delta_{n-1}}(z_{n-1})
    \scalar^{\out}_{\Delta_n}(z_n)
    \sim
    \scalar^{\out}_{\Delta_{n-1}+\Delta_n-2}(z_n)
    \, .
\end{align}
Then we take the OPE of this exchange operator with $\scalar^{\out}_{\Delta_{n-2}}$, which gives
\begin{align}
    \scalar^{\out}_{\Delta_{n-2}}(z_{n-2})
    \scalar^{\out}_{\Delta_{n-1}+\Delta_n-2}(z_n)
    \sim
    \scalar^{\out}_{\Delta_{n-2}+\Delta_{n-1}+\Delta_n-4}(z_n)
    \, .
\end{align}
Continuing in this way, the outgoing cluster $\scalar^{\out}_{\Delta_3}\cdots \scalar^{\out}_{\Delta_n}$ can be replaced, in the corresponding multiple-collinear limit, by an exchange outgoing massless scalar primary $\scalar^{\out}_{\Delta_{3\cdots n}-2n+6}$.
Thus the $n$-point function reduces, in this iterated OPE limit, to an effective three-point function of the form $\vev{\scalar^{\inn}_{\Delta_1}\scalar^{\inn}_{\Delta_2}\scalar^{\out}_{\Delta_{3\cdots n}-2n+6}}$.
This is precisely the type of three-point function considered above. If the $12$ OPE contained only the ordinary Mellin-basis operator $\scalar^{\inn}_{\Delta_{12}-2}$, then the remaining two-point function with the exchange outgoing operator would be contact-supported on the celestial sphere and would vanish when the two OPE centres are separated. Therefore the same separated-OPE consistency argument implies that the $12$ OPE must also contain the shadow-basis operator $\shadow{\scalar^{\inn}_{\Delta_{12}-2}}$.

This is also consistent with the overall scaling support. A tree-level massless $\phi^3$ $n$-point celestial amplitude is supported on $\delta(\Delta_{1\cdots n}-2n+2)$. On this support one has
\begin{align}
    (\Delta_{12}-2)
    +
    (\Delta_{3\cdots n}-2n+6)
    =
    2
    \, .
\end{align}
Thus the exchange outgoing operator $\scalar^{\out}_{\Delta_{3\cdots n}-2n+6}$ has precisely the dimension needed to pair with the shadow of the $12$-channel operator through a standard non-contact two-point function. Hence the shadow exchange found in the three-point analysis is not special to three-point functions; it is already present in any higher-point scalar correlator after reducing one cluster by ordinary collinear OPEs.

Therefore the ordinary massless OPE should be completed as
\begin{align}
    \scalar^{\inn}_{\Delta_1}(z_1)
    \scalar^{\inn}_{\Delta_2}(z_2)
    \sim
    \frac{1}{|z_{1,2}|^2}\coOPE_{\Delta_{12}-2}\,
    \scalar^{\inn}_{\Delta_{12}-2}(z_2)
    +
    \frac{1}{|z_{1,2}|^{2\Delta_{12}-4}}\wave{\coOPE}_{4-\Delta_{12}}\,
    \wave{\scalar}^{\inn}_{4-\Delta_{12}}(z_2)
    \, .
\end{align}

%%%%%%%%%%%%%%%%%%%%%%%%%%%%%%%%%%%%%%%%%%%%%%%%%%%%%%%%%%%%%%%%%%%%%%%%%
\subsection{Shadow-basis states and the boundary Hilbert space}
%%%%%%%%%%%%%%%%%%%%%%%%%%%%%%%%%%%%%%%%%%%%%%%%%%%%%%%%%%%%%%%%%%%%%%%%%

The appearance of the shadow-basis operator in the OPE raises a natural question. If the OPE contains an operator $\shadow{\scalar_{\Delta}}$, then one can construct the state
\begin{align}
    \shadow{\scalar_{\Delta}}(0)\vac
\end{align}
in the boundary Hilbert space, where $\vac$ is the boundary vacuum. Should this state be regarded as an independent state in the boundary Hilbert space of the celestial theory, or can it be written as a convergent linear combination of the Mellin-basis state $\scalar_{\Delta}(0)\vac$ and its descendants?

There are two distinct notions of independence which should be separated. The shadow-basis state does not represent a new bulk particle. Both $\scalar_{\Delta}$ and $\shadow{\scalar_{\Delta}}$ are obtained from the same four-dimensional massless scalar one-particle representation, but with different choices of conformal primary basis on the celestial sphere. As a basis for the same bulk one-particle representation, the shadow basis is related to the Mellin basis by a linear integral transform and is therefore not independent. In this sense, the shadow transform only changes the celestial representative of the same bulk state and does not add a new bulk degree of freedom.

On the other hand, as a local state in the boundary CFT, $\shadow{\scalar_{\Delta}}(0)\vac$ is not contained in the conformal family generated by $\scalar_{\Delta}(0)\vac$. To see this, suppose that $\shadow{\scalar_{\Delta}}(0)\vac$ could be written as a convergent linear combination of ordinary descendants,
\begin{align}
    \shadow{\scalar_{\Delta}}(0)\vac
    =
    \sum_{m,n\geq 0}
    a_{mn}\,
    \partial^m\bar{\partial}^n
    \scalar_{\Delta}(0)\vac
    \, ,
\end{align}
where $a_{mn}$ are constants.
Let $h=\bar h=\frac{\Delta}{2}$ be the weights of the scalar Mellin primary. The descendants of $\scalar_{\Delta}(0)\vac$ have weights $(h+m,\bar h+n)$ with $m,n\in\mathbb Z_{\geq0}$. By contrast, the shadow primary has weights $(1-h,1-\bar h)$. If the shadow state were contained in the same local descendant module, one would need $1-h=h+m$ and $1-\bar h=\bar h+n$ for some non-negative integers $m,n$.
Thus, except possibly at special values $h=(1-m)/2$ and $\bar h=(1-n)/2$ with $m,n\in\mathbb Z_{\geq0}$, the shadow state $\shadow{\scalar_{\Delta}}(0)\vac$ cannot be written as a convergent linear combination of $\scalar_{\Delta}(0)\vac$ and its descendants.

The same conclusion is reflected in the definition of the shadow transform. For a scalar primary we have
\begin{align}
    \shadow{\scalar_{\Delta}}(0)\vac
    =
    \int d^2 z'\,
    \frac{1}{|z'|^{2(2-\Delta)}}\,
    \scalar_{\Delta}(z')\vac
    \, .
\end{align}
Using translation covariance,
\begin{align}
    \scalar_{\Delta}(z)\vac
    =
    e^{zL_{-1}+\zb\bar L_{-1}}
    \scalar_{\Delta}(0)\vac
    \, ,
\end{align}
one might try to expand the exponential and interpret the shadow state as a linear combination of descendants. This gives the formal expression
\begin{align}
    \shadow{\scalar_{\Delta}}(0)\vac
    =
    \sum_{m,n\geq0}
    \frac{L_{-1}^{m}\bar L_{-1}^{n}}{m!\,n!}
    \scalar_{\Delta}(0)\vac\,
    I_{m,n}(\Delta)
    \, ,
\end{align}
with
\begin{align}
    I_{m,n}(\Delta)
    =
    \int d^2 z'\,
    (z')^m (\zb')^n
    |z'|^{-2(2-\Delta)}
    \, .
\end{align}
However, these coefficients are not ordinary convergent integrals. Writing $z'=re^{i\theta}$, one finds
\begin{align}
    I_{m,n}(\Delta)
    =
    \int_0^\infty dr\,
    r^{m+n+2\Delta-3}
    \int_0^{2\pi}d\theta\,
    e^{i(m-n)\theta}
    \, .
\end{align}
The angular integral vanishes unless $m=n$. For $m=n$, the radial integral becomes
\begin{align}
    I_{m,m}(\Delta)
    =
    2\pi
    \int_0^\infty dr\,
    r^{2m+2\Delta-3}
    \, ,
\end{align}
which is not a convergent integral in the ordinary sense.%
\footnote{
    With a distributional or regularized prescription, such radial integrals may be interpreted in terms of complex-support delta functions \cite{Gelfand1,Gelfand2,Morimoto1,Donnay:2020guq,Liu:2024lbs,Liu:2025dhh,Liu:2025voe,Liu:2026toc}.
    The condition for the radial measure to take the scale-invariant form is $\Delta=1-m$ for scalar primaries. This does not give a convergent descendant expansion for generic $\Delta$.
}
Consequently, $\shadow{\scalar_{\Delta}}(0)\vac$ should not be identified with a convergent linear combination of $\scalar_{\Delta}(0)\vac$ and its descendants.

At first sight, this may seem to be in tension with the statement that the Mellin and shadow bases are not independent in the bulk. The point is that two different notions of representation are being compared. In the bulk description, the Mellin conformal primary basis with fixed $\Delta$ furnishes a principal-series representation of the Lorentz group. The shadow transform acts within the same principal-series representation and gives an equivalent realization of it. Thus, at the level of the bulk one-particle Hilbert space, the Mellin and shadow bases are related by an isomorphism and should not be counted as independent bulk degrees of freedom.

The situation is different after the representation is realized as a Verma module in the boundary CCFT. The principal-series representation and the corresponding boundary module share the same transformation properties under the complexified conformal algebra, so the bulk conformal primary basis can be identified with boundary primary states \cite{Liu:2026toc}. However, the Verma module generated by $\scalar_{\Delta}(0)\vac$ contains only the descendants obtained by acting with local translation generators. The shadow primary has weights $(1-h,1-\bar h)$, and is not generated from $\scalar_{\Delta}(0)\vac$ by such local descendant operations. Thus, \textit{the Mellin and shadow bases are isomorphic descriptions of the same bulk principal-series representation, but they correspond to distinct local Verma modules once realized in the boundary CCFT.}

%%%%%%%%%%%%%%%%%%%%%%%%%%%%%%%%%%%%%%%%%%%%%%%%%%%%%%%%%%%%%
\subsection{OPE coefficients of exchange shadow operators}
%%%%%%%%%%%%%%%%%%%%%%%%%%%%%%%%%%%%%%%%%%%%%%%%%%%%%%%%%%%%%

We now determine the coefficient of the shadow term. The important point is that $\wave{\coOPE}_{4-\Delta_{12}}$ is not an independent OPE datum, but is related to the ordinary collinear OPE coefficient by the universal shadow factor. To see this, consider the scalar three-point functions
\begin{align}
    &\vev{
        \scalar^{\inn}_{\Delta_1}
        \scalar^{\inn}_{\Delta_2}
        \scalar^{\out}_{\Delta_3}
    }
    =
    \coThree(\scalar^{\inn}_{\Delta_1}
        \scalar^{\inn}_{\Delta_2}
    \scalar^{\out}_{\Delta_3})
    \vevv{
        \scalar^{\inn}_{\Delta_1}
        \scalar^{\inn}_{\Delta_2}
        \scalar^{\out}_{\Delta_3}
    }
    \, ,\\
    &\vev{
        \scalar^{\inn}_{\Delta_1}
        \scalar^{\inn}_{\Delta_2}
        \wave{\scalar}^{\out}_{2-\Delta_3}
    }
    =
    \coThree(\scalar^{\inn}_{\Delta_1}
        \scalar^{\inn}_{\Delta_2}
    \wave{\scalar}^{\out}_{2-\Delta_3})
    \vevv{\scalar^{\inn}_{\Delta_1}
        \scalar^{\inn}_{\Delta_2}
        \wave{\scalar}^{\out}_{2-\Delta_3}
    }
    \, ,
\end{align}
where the three-point coefficients of these two three-point functions are related by
\begin{align}
    \coThree(\scalar^{\inn}_{\Delta_1}
        \scalar^{\inn}_{\Delta_2}
    \wave{\scalar}^{\out}_{2-\Delta_3})
    =
    \coShadow_{\Delta_3}^{\Delta_1,\Delta_2}
    \coThree(\scalar^{\inn}_{\Delta_1}
        \scalar^{\inn}_{\Delta_2}
    \scalar^{\out}_{\Delta_3})
    \, .
\end{align}
Here $\coShadow_{\Delta_3}^{\Delta_1,\Delta_2}$, given in \eqref{eq:shadow_factor}, is the universal shadow factor obtained by applying the two-dimensional shadow transform to the third leg.

Now, take the OPE limit $z_1\to z_2$, while keeping $z_2\neq z_3$. In the separated region, the OPE coefficient can be extracted by dividing the three-point coefficient by the coefficient of the remaining mixed two-point function. For the ordinary Mellin-basis exchange and the shadow-basis exchange, respectively, we have
\begin{align}
    \coOPE_{\Delta_{12}-2}=\frac{\coThree(\scalar^{\inn}_{\Delta_1}\scalar^{\inn}_{\Delta_2}\wave{\scalar}^{\out}_{\Delta_{12}-2})}{\coTwo(\scalar^{\inn}_{\Delta_{12}-2}\wave{\scalar}^{\out}_{\Delta_{12}-2})},\hspace{1cm}
    \wave{\coOPE}_{4-\Delta_{12}}=\frac{\coThree(\scalar^{\inn}_{\Delta_1}\scalar^{\inn}_{\Delta_2}\scalar^{\out}_{4-\Delta_{12}})}{\coTwo(\wave{\scalar}^{\inn}_{4-\Delta_{12}}\scalar^{\out}_{4-\Delta_{12}})}
    \, .
\end{align}
Since $\coTwo(
    \scalar^{\inn}_{\Delta_{12}-2}
    \wave{\scalar}^{\out}_{\Delta_{12}-2}
) =
\coTwo(
    \wave{\scalar}^{\inn}_{4-\Delta_{12}}
    \scalar^{\out}_{4-\Delta_{12}}
)$, the two extractions differ only by the shadow factor relating the two three-point functions.\footnote{
    Both mixed two-point functions are obtained by shadow transforming one leg of
    the same contact-supported Mellin two-point function
    $\vev{\scalar^{\inn}_{\Delta_{12}-2}(z_2)
    \scalar^{\out}_{4-\Delta_{12}}(z_3)}
    \propto \delta^{(2)}(z_{2,3})$. With the normalization
    \eqref{eq: shadow transform}, shadowing the incoming or outgoing leg gives
    the same mixed two-point coefficient.
}  Therefore the OPE coefficients obey
\begin{align}
    \coOPE_{\Delta_{12}-2}
    =
    \coShadow_{4-\Delta_{12}}^{\Delta_1,\Delta_2}\,
    \wave{\coOPE}_{4-\Delta_{12}}
    \, .
\end{align}
Since the ordinary collinear OPE coefficient in massless $\phi^3$ theory is
\begin{align}
    \coOPE_{\Delta_{12}-2}=\frac{1}{4}\bt*{\Delta_1-1,\Delta_2-1}
    \, ,
\end{align}
we find
\begin{align}
    \wave{\coOPE}_{4-\Delta_{12}}=\frac{1}{4\pi}\mg*{2-\Delta_1, 2-\Delta_2}{3-\Delta_{12}}
    \, .
\end{align}

%%%%%%%%%%%%%%%%%%%%%%%%%%%%%%%%%%%%%%%%%%%%%%%%%%%%%%%%%%%%%
\subsection{Gluons and gravitons}
%%%%%%%%%%%%%%%%%%%%%%%%%%%%%%%%%%%%%%%%%%%%%%%%%%%%%%%%%%%%%

The preceding discussion was carried out for scalar primaries, but the mechanism is not special to scalar fields. The same reasoning applies to massless particles with spin, in particular to gluons and gravitons. The only additional ingredient is that the exchanged operator carries spin, and therefore the shadow transform acts both on the conformal dimension and on the two-dimensional spin.

Let ${\cal O}_{\Delta,J}$ denote a massless celestial primary of spin $J$. The ordinary collinear OPE has the schematic form
\begin{align}
    {\cal O}_{\Delta_1,J_1}(z_1)\,
    {\cal O}_{\Delta_2,J_2}(z_2)
    \sim
    \coOPE_{\Delta,J}\,
    {\cal O}_{\Delta,J}(z_2)
    \, ,
\end{align}
where $\Delta$, $J$, and $\coOPE_{\Delta,J}$ are fixed by the corresponding collinear splitting function or by the asymptotic symmetries. If this ordinary Mellin-basis operator were the only exchanged operator, then its two-point contraction with an operator in a separated OPE cluster would again be contact-supported on the celestial sphere. Therefore it cannot by itself reproduce the non-contact power-law behavior required by a separated OPE limit of a regular celestial correlator.

The same separated-OPE consistency argument then implies that the OPE must also contain the shadow-basis representative of the exchanged particle $\wave{{\cal O}}_{2-\Delta,-J}$. The coefficient of the shadow term is again fixed by the ordinary OPE coefficient and by the universal shadow transform of the exchanged leg. If the ordinary OPE coefficient is denoted by $\coOPE_{\Delta,J}$, then
\begin{align}\label{eq:shadow_and_mellin_ope_coefficient}
    \coOPE_{\Delta,J}
    =
    \coShadow_{2-\Delta,-J}^{\Delta_1,J_1,\Delta_2,J_2}
    \wave{\coOPE}_{2-\Delta,-J}
    \, .
\end{align}
Here $\coShadow_{2-\Delta,-J}^{\Delta_1,J_1,\Delta_2,J_2}$, given in \eqref{eq:shadow_factor}, is the universal shadow factor obtained by applying the two-dimensional shadow transform to the exchanged operator in the corresponding three-point structure.

In summary, the shadow-completed scalar OPE is
\begin{align}
    \label{eq:OPE_scalar}
    \scalar^{\inn}_{\Delta_1}
    \scalar^{\inn}_{\Delta_2}
    \sim
    \frac{1}{4}\bt*{\Delta_1-1,\Delta_2-1}
    \scalar^{\inn}_{\Delta_{12}-2}
    +
    \frac{1}{4\pi}\mg*{2-\Delta_1, 2-\Delta_2}{3-\Delta_{12}}
    \wave{\scalar}^{\inn}_{4-\Delta_{12}}
    \, .
\end{align}
For gluons, the corresponding shadow-completed OPE is
\begin{align}\label{eq:OPE_gluon}
    \gluon^{\inn,a}_{\Delta_1,+}
    \gluon^{\inn,b}_{\Delta_2,+}
    \sim
    if^{abc}
    \bt*{\Delta_1-1,\Delta_2-1}
    \gluon^{\inn,c}_{\Delta_{12}-1,+}
    -
    \frac{if^{abc}}{\pi}
    \mg*{1-\Delta_1,1-\Delta_2}{1-\Delta_{12}}
    \wave{\gluon}^{\inn,c}_{3-\Delta_{12},-}
    \, .
\end{align}
For gravitons, one similarly obtains
\begin{align}\label{eq:OPE_graviton}
    \graviton^{\inn}_{\Delta_1,+}
    \graviton^{\inn}_{\Delta_2,+}
    \sim
    -\half
    \bt*{\Delta_1-1,\Delta_2-1}
    \graviton^{\inn}_{\Delta_{12},+}
    -
    \frac{1}{2\pi}
    \mg*{\Delta_1,\Delta_2}{-\Delta_{12}-1}
    \wave{\graviton}^{\inn}_{2-\Delta_{12},-}
    \, .
\end{align}
In the above OPEs we have suppressed the coordinate dependence for notational simplicity.

These OPEs are obtained from the same OPE-consistency logic. Whenever the ordinary Mellin-basis exchange gives only a contact-supported two-point function in a separated OPE limit of a regular celestial correlator, consistency with the non-contact conformal two-point structure requires the shadow-basis representative of the same exchanged massless particle.

%%%%%%%%%%%%%%%%%%%%%%%%%%%%%%%%%%%%%%%%%%%%%%%%%%%%%%%%%%%%%
\subsection{From shadow-completed OPEs to conformal partial waves}
%%%%%%%%%%%%%%%%%%%%%%%%%%%%%%%%%%%%%%%%%%%%%%%%%%%%%%%%%%%%%

The shadow-completed OPE has a direct implication at the level of four-point functions. Suppose that the OPE of two operators contains both a Mellin-basis exchange operator and its shadow-basis representative,
\begin{align}
    \op_1\op_2
    \sim
    \coOPE_{\Delta,J}\,
    \op_{\Delta,J}
    +
    \wave{\coOPE}_{\widetilde\Delta,\widetilde J}\,
    \wave{\op}_{\widetilde\Delta,\widetilde J}
    \, .
\end{align}
Consequently, the four-point function $\vev{\op_1\op_2\op_3\op_4}$ receives two conformal-block contributions: one from the exchange of $(\Delta,J)$, and one from the exchange of its shadow $(\widetilde\Delta,\widetilde J)=(2-\Delta,-J)$. Schematically,
\begin{align}
    \vev{\op_1\op_2\op_3\op_4}
    &=
    \coOPE_{\Delta,J}\,
    \coThree(\op_{\Delta,J}\op_3\op_4)\,
    G_{\Delta,J}(z_i)
    +
    \wave{\coOPE}_{\widetilde\Delta,\widetilde J}\,
    \coThree(\wave{\op}_{\widetilde\Delta,\widetilde J}\op_3\op_4)\,
    G_{\widetilde{\Delta},\widetilde{J}}(z_i)
    +\cdots
    \, .
\end{align}
These two terms are related by the shadow transform. The corresponding three-point coefficients obey
\begin{align}
    \coThree(\wave{\op}_{\widetilde\Delta,\widetilde J}\op_3\op_4)
    =
    \coShadow_{\Delta,J}^{\Delta_3,J_3,\Delta_4,J_4}\,
    \coThree(\op_{\Delta,J}\op_3\op_4)
    \, .
\end{align}
Equivalently, using the inverse shadow transform, we have
\begin{align}
    \coThree(\op_{\Delta,J}\op_3\op_4)
    =
    N_{\Delta,J}\,
    \coShadow_{\widetilde\Delta,\widetilde J}^{\Delta_3,J_3,\Delta_4,J_4}\,
    \coThree(\wave{\op}_{\widetilde\Delta,\widetilde J}\op_3\op_4)
    \, ,
\end{align}
where $N_{\Delta,J}$ is the normalization appearing in \eqref{eq:inverse_shadow_transform}. On the other hand, the shadow-completed OPE implies \eqref{eq:shadow_and_mellin_ope_coefficient}. Substituting these relations into the four-point contribution yields
\begin{align}
    \vev{\op_1\op_2\op_3\op_4}
    &=
    N_{\Delta,J}\,
    \coOPE_{\Delta,J}\,
    \coThree(\wave{\op}_{\widetilde\Delta,\widetilde J}\op_3\op_4)
    % \nn
    % \\
    % &\quad \times
    \left(
        \coShadow_{\widetilde\Delta,\widetilde J}^{\Delta_3,J_3,\Delta_4,J_4}\,
        G_{\Delta,J}(z_i)
        +
        \coShadow_{\Delta,J}^{\Delta_1,J_1,\Delta_2,J_2}\,
        G_{\widetilde\Delta,\widetilde J}(z_i)
    \right)
    +\cdots
    \, .
\end{align}
The expression in brackets is the conformal partial wave built from the exchanged representation $(\Delta,J)$:
\begin{align}
    \Psi_{\Delta,J}(z_i)
    \eqq
    \coShadow_{\widetilde\Delta,\widetilde J}^{\Delta_3,J_3,\Delta_4,J_4}\,
    G_{\Delta,J}(z_i)
    +
    \coShadow_{\Delta,J}^{\Delta_1,J_1,\Delta_2,J_2}\,
    G_{\widetilde\Delta,\widetilde J}(z_i)
    \, .
\end{align}
Therefore,
\begin{align}
    \vev{\op_1\op_2\op_3\op_4}
    =
    N_{\Delta,J}\,
    \coOPE_{\Delta,J}\,
    \coThree(\wave{\op}_{\widetilde\Delta,\widetilde J}\op_3\op_4)\,
    \Psi_{\Delta,J}(z_i)
    +\cdots
    \, .
\end{align}
Thus, \textit{once the OPE contains both the Mellin-basis exchange and its shadow
    exchange with the shadow-related OPE coefficients, the two conformal blocks
assemble into a single conformal partial wave.}

It is useful to distinguish this from the usual conformal partial wave expansion. In a conformal partial wave expansion, $\Delta$ is an integration variable along the principal series,
\begin{align}
    \vev{\op_1\op_2\op_3\op_4}
    =
    \sum_J
    \intrange{\frac{d\Delta}{2\pi i}}{1}{1+i\oo}
    \rho_{\Delta,J}\,
    \Psi_{\Delta,J}(z_i)
    \, .
\end{align}
By contrast, in the OPE analysis above, $\Delta$ is not an integration variable. Rather, it is fixed by the exchanged operator that appears in the collinear OPE. For example, in massless $\phi^3$ theory, the exchanged Mellin-basis operator in the $12$ OPE has fixed dimension
\begin{align}
    \Delta=\Delta_{12}-2
    \, .
\end{align}
The shadow-completed OPE therefore directly gives two fixed conformal blocks,
\begin{align}
    G_{\Delta_{12}-2}(z_i)
    \, ,
    \qquad
    G_{2-(\Delta_{12}-2)}(z_i)
    \, .
\end{align}
Equivalently, if one starts from the conformal partial wave expansion and deforms the $\Delta$-contour, these two blocks arise as residues at the collinear pole $\Delta=\Delta_{12}-2$ and its shadow pole $\Delta=2-(\Delta_{12}-2)$. The shadow relation between the two OPE coefficients ensures that the two residues have the relative normalization needed to recombine into a single conformal partial wave $\Psi_{\Delta_{12}-2}$.

%%%%%%%%%%%%%%%%%%%%%%%%%%%%%%%%%%%%%%%%%%%%%%%%%%%%%%%%%%%%%%%%
\section{Tree-level exchange in celestial holography} \label{sec:tree_level_exchange}
%%%%%%%%%%%%%%%%%%%%%%%%%%%%%%%%%%%%%%%%%%%%%%%%%%%%%%%%%%%%%%%%
Our goal in this section is to extract the shadow OPE of two incoming massless scalar particles from regular celestial amplitudes. We work with \TwoToN massless scalar scattering and show that the tree-level exchange contribution universally contains a shadow exchange operator in the OPE of the two incoming particles.

The analysis follows the regular celestial amplitude framework of \cite{Liu:2025voe}. Although a regular celestial amplitude is defined by mass-regularizing all external particles, a specific OPE channel can be extracted by regularizing only the particles relevant to that channel. This reduced regularization was used in \cite{Liu:2025voe} to recover the collinear OPE of two massless particles, in agreement with the collinear splitting analysis of \cite{Fan:2019emx,Pate:2019lpp}.

We apply the same strategy here. With single-particle mass regularization, the OPE limit isolates the shadow exchange contribution, whose OPE coefficient agrees with the structures found in \cite{Himwich:2025bza}.

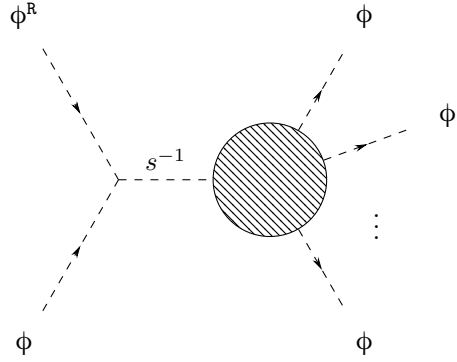
\includeImage{celestial-2-n}{Tree-level \TwoToN exchange contribution used to extract the $12$-channel OPE from the regular celestial amplitude. The first incoming scalar is mass-regularized, the internal line denotes the $s$-channel propagator, and the shaded blob represents the remaining \OneToN subamplitude.}

%%%%%%%%%%%%%%%%%%%%%%%%%%%%%%%%%%%%%%%%%%%%%
\subsection{Regular celestial amplitude and split representation}
%%%%%%%%%%%%%%%%%%%%%%%%%%%%%%%%%%%%%%%%%%%%%

\paragraph{Regularized conformal basis.}

We briefly recall the regularized conformal basis for scalar particles \cite{Liu:2025voe},
\begin{align}
    \label{eq: regularized conformal basis}
    \scalar^{\innout,\reg}_{\Delta}
    &=
    \thalf \mlimit_{\Delta}^{m} \LR{
        \scalar^{\innout,m}_{\Delta}
        -
        \scalar^{\tacpm,im}_{\Delta}
    }
    \, ,
    \\
    \wave{\scalar}^{\innout,\reg}_{\Delta}
    &=
    \thalf \LR{
        \scalar^{\innout,m}_{\Delta}
        +
        \scalar^{\tacmp,im}_{\Delta}
    }
    \, .
\end{align}
Here $\scalar^{\innout,m}_{\Delta}$ and $\scalar^{\tacpm,im}_{\Delta}$ are the massive and tachyonic conformal bases, and their explicit expressions can be found in \cite{Liu:2025voe}. The massless limit factor is
\begin{equation}
    \label{eq: massless limit factor}
    \mlimit_{\Delta}^{m}
    =
    2^{2-2 \Delta}
    m^{2 \Delta -2}
    \pi^{-1}
    (\Delta-1)
    \, .
\end{equation}
Replacing each ordinary massless conformal basis element by its regularized counterpart then defines the corresponding regular celestial amplitude.

\paragraph{Mass-dependence of celestial amplitudes.}
For the \OneToN scattering process with one massive particle, the mass dependence of the celestial amplitude can be extracted by the dimensional analysis,
\begin{equation}
    \label{eq: 1toN process - celestial amplitude mass dependence}
    \vev{\scalar^{\inn,m}_{\Delta}(z)\rest}
    =
    m^{\delta-\Delta-2n+4}
    f^{\inn}_{\Delta,\Delta_{i}}(z,z_{i})
    \, ,
\end{equation}
where $\delta = \sum \Delta_{i}$ is the total dimension of the other operators.
Then the \OneToN regular celestial amplitude is
\begin{equation}
    \label{eq: 1toN process - regular celestial amplitude}
    \vev{\wave{\scalar}^{\inn,\reg}_{\Delta}(z)\rest}
    =
    \thalf
    \deltaK*{\delta-\Delta-2n+4}
    f^{\inn}_{\Delta,\Delta_{i}}(z,z_{i})
    \, .
\end{equation}
The factor $\half$ comes from \eqref{eq: regularized conformal basis}, and there are no tachyonic contributions due to momentum conservation.

\paragraph{Split representation.}
For a function of squared momentum $G(p^{2})$, the split representation is
\begin{equation}
    \label{eq: split representation}
    \begin{aligned}
    &
    G(p^{2})
    =
    \int[d\Delta dm]
    \int d^{2}z
    \LRb{
        G(-m^2)
        \LRa{
            \scalar^{\inn,m}_{\Delta}
            \scalar^{\out,m}_{\DeltaS}
            +
            \scalar^{\out,m}_{\Delta}
            \scalar^{\inn,m}_{\DeltaS}
        }
        +
        G(m^2)
        \LRa{
            \scalar^{\tacp,im}_{\Delta}\scalar^{\tacm,im}_{\DeltaS}
            +
            \scalar^{\tacm,im}_{\Delta}\scalar^{\tacp,im}_{\DeltaS}
        }
    }
    +
    \rest
    \, .
    \end{aligned}
\end{equation}
Here the two terms correspond to massive and tachyonic momenta respectively, and $\rest$ denotes discrete series contributions of tachyonic momentum; see \cite{Chang:2023ttm}.
The Plancherel measure is
\begin{equation}
    \intt{[d\Delta dm]}
    =
    \intrange{dm}{0}{\oo}
    \intrange{\frac{d\Delta}{2\pi i} \mu^{m}_{\Delta}}{\halfdim}{\halfdim+i\oo}
    \, ,
    \textInMath{where}
    \mu^{m}_{\Delta}
    =
    2^{d}
    \pi^{-d}
    m
    \mg*{
        \Delta,\DeltaS
    }{
        \Delta-\halfdim,
        \DeltaS-\halfdim
    }
    \, .
\end{equation}
%%%%%%%%%%%%%%%%%%%%%%%%%%%%%%%%%%%%%%%%%%%%%%%%%%%%%%%%%%%%%%%%%%%%%%%%%%%%%%
\subsection{Shadow exchange from scalar \TwoToN tree-level scattering}\label{sec: 2toN process - shadow exchange}
%%%%%%%%%%%%%%%%%%%%%%%%%%%%%%%%%%%%%%%%%%%%%%%%%%%%%%%%%%%%%%%%%%%%%%%%%%%%%%%
For a \TwoToN process, we take the mass regularization to the first particle,
\begin{equation}
    \label{eq: 2toN process - regular celestial amplitude}
    \vev{\scalar^{\inn,\reg}_{\Delta_1}(z_1)\scalar^{\inn}_{\Delta_2}(z_2)\rest}
    =
    \lim_{m_{1}\to 0}
    \thalf
    \mlimit_{\Delta_{1}}^{m_{1}}
    \LR{
        \vev{\scalar^{\inn,m_{1}}_{\Delta_1}(z_1)\scalar^{\inn}_{\Delta_2}(z_2)\rest}
        -
        \vev{\scalar^{\tacp,im_{1}}_{\Delta_1}(z_1)\scalar^{\inn}_{\Delta_2}(z_2)\rest}
    }
    \, .
\end{equation}
We consider the following tree-level contributions to the regular celestial amplitude \eqref{eq: 2toN process - regular celestial amplitude}%
\footnote{
    We have suppressed the $\ii$-factor in the propagator and the coupling constant for simplicity.
}
\begin{equation}
    \label{eq: 2toN process - scattering amplitude - leading contribution}
    \cT_{\TwoToN}(q_1,q_2,\set{q_{i}})
    =
    \frac{1}{(q_1+q_2)^2}
    \cT_{\OneToN}(q_1+q_2,\set{q_{i}})
    \, ,
\end{equation}
where $q_{i}$ denotes the momenta of the rest $n$-particles.
Inserting the split representation \eqref{eq: split representation} into \eqref{eq: 2toN process - regular celestial amplitude}, each exchange term from \eqref{eq: split representation} is rewritten into an integral over a product of \TwoToOne and \OneToN celestial amplitudes.
For the tachyonic continuous and discrete ones from \eqref{eq: split representation}, the corresponding \OneToN amplitudes is strictly forbidden by the momentum conservation, hence we only need to consider the massive \OneToN contributions to \eqref{eq: 2toN process - regular celestial amplitude}.

\paragraph{Tachyonic external particle.}
We first consider the second term in \eqref{eq: 2toN process - regular celestial amplitude},
\begin{equation}
    \begin{aligned}
    \vev{\scalar^{\tacp,im_{1}}_{\Delta_1}(z_1)\scalar^{\inn}_{\Delta_2}(z_2)\rest}
    &=
    \intt{[d\Delta_{0} dm_{0}]}
    m_{0}^{-2}
    \intt{d^{2}z_{0}}
    \vev{\scalar^{\tacp,im_{1}}_{\Delta_1}(z_1)\scalar^{\inn}_{\Delta_2}(z_2)\scalar^{\out,m_{0}}_{\DeltaS_0}(z_0)}
    \vev{\scalar^{\inn,m_{0}}_{\Delta_{0}}(z_0)\rest}
    \, ,
    \end{aligned}
\end{equation}
where $m_{0}^{-2}$ is from the propagator and $\vev{\scalar^{\inn,m_{0}}_{\Delta_{0}}\rest}$ is defined by the decay amplitude $\cT_{\OneToN}$ in \eqref{eq: 2toN process - scattering amplitude - leading contribution}.

We first consider the $z_{0}$-integral.
The three-point correlator takes the standard form, see \cite[Appendix D]{Liu:2025voe}:
\begin{equation}
    \vev{\scalar^{\tacp,im_{1}}_{\Delta_1}\scalar^{\inn}_{\Delta_2}\scalar^{\out,m_{0}}_{\DeltaS_0}}
    =
    \coThree(\scalar^{\tacp,im_{1}}_{\Delta_1}\scalar^{\inn}_{\Delta_2}\scalar^{\out,m_{0}}_{\DeltaS_0})
    \vevv{\scalar_{\Delta_{1}}\scalar_{\Delta_{2}}\scalar_{\DeltaS_{0}}}
    \, ,
\end{equation}
where the coefficient is schematically,
\begin{equation}
    \label{eq: 3pt coefficient - TOM}
    \begin{aligned}
    &
    \coThree(\scalar^{\tacp,im_{1}}_{\Delta_1}\scalar^{\inn}_{\Delta_2}\scalar^{\out,m_{0}}_{\DeltaS_0})
    =
    (\rest)
    \xx
    \Fpq{2}{1}{\half \Delta_{02,1}, \half (\Delta_{2,01}+2)}{2-\Delta_1}{-\frac{m_1^2}{m_0^2}}
    \, .
    \end{aligned}
\end{equation}
Then the $z_{0}$-integral takes the form of \eqref{eq: conformal integral for OPE - EAdS}, and in the OPE limit $z_{1} \to z_{2}$ can be estimated by \eqref{eq: conformal integral for OPE - EAdS - OPE limit}.
There are two leading terms from the estimation \eqref{eq: conformal integral for OPE - EAdS - OPE limit}, but after changing the variable $\Delta_{0}\to\DeltaS_{0}$ for the second, it turns out that the two terms are identical with different integration contours, which can be combined together into the principal series from $1-i\oo$ to $1+i\oo$.
In the OPE limit $r = \abs{z_{1,2}} \to 0$, the remaining leading term is
\begin{equation}
    \begin{aligned}
    \vev{\scalar^{\tacp,im_{1}}_{\Delta_1}(z_1)\scalar^{\inn}_{\Delta_2}(z_2)\rest}
    &=
    r^{\Delta_{0,12}}
    \xx
    \intrange{\frac{d\Delta_{0}}{2\pi i}}{1-i\oo}{1+i\oo}
    \intrange{dm_{0}}{0}{\oo}
    \rho^{m_{0}}_{\Delta_{0}}
    \vev{\scalar^{\inn,m_{0}}_{\Delta_{0}}(z_{2})\rest}
    +
    \SmallO(r^{\Delta_{0,12}})
    \, ,
    \end{aligned}
\end{equation}
where the density is
\begin{equation}
    \rho^{m_{0}}_{\Delta_{0}}
    =
    m_{0}^{-2}
    \mu^{m_{0}}_{\Delta_{0}}
    \coThree(\scalar^{\tacp,im_{1}}_{\Delta_1}\scalar^{\inn}_{\Delta_2}\scalar^{\out,m_{0}}_{\DeltaS_0})
    I_{1}
    \, .
\end{equation}

Now we insert the mass dependence of the \OneToN celestial amplitude \eqref{eq: 1toN process - celestial amplitude mass dependence}, then the $m_{0}$-integral can be evaluated by the Mellin-Barnes representation of the hypergeometric function together with the Mellin-Barnes lemma, giving
\begin{align}
    \label{eq: 2toN process - OPE limit - TOM contribution}
    &
    \vev{\scalar^{\tacp,im_{1}}_{\Delta_1}(z_1)\scalar^{\inn}_{\Delta_2}(z_2)\rest}
    =
    -r^{\Delta_{0,12}}
    \xx
    2^{\Delta_{1,02}-1}
    m_1^{\Delta_{2,1}+\delta -2 n+2}
    \mg*{
        1-\Delta_1,
        \half (2 n -\Delta_{12}-\delta)
    }{
        \half (\Delta_{2,1}+\delta -2 n+4)
    }
    \\
    & \peq
    \xx
    \intrange{\frac{d\Delta_{0}}{2\pi i}}{1-i\oo}{1+i\oo}
    f^{\inn}_{\Delta_0,\Delta_{i}}(z_{2},z_{i})
    \mg*{
        \half (-\Delta_0+\delta -2 n+4)
        ,
        \half (\Delta_0+\delta -2 n+2)
        ,
        \half \Delta_{01,2},
        \half \Delta_{02,1}
    }{
        \Delta_0-1,
        \half (\Delta_{0,12}+2),
        \half (4-\Delta_{012})
    }
    +
    \SmallO(r^{\Delta_{0,12}})
    \, .
    \nn
\end{align}

\paragraph{Massive external particle.}
Similarly, for the first term in \eqref{eq: 2toN process - regular celestial amplitude}, the leading contribution in the OPE limit is
\begin{align}
    &
    \vev{\scalar^{\inn,m_{1}}_{\Delta_1}(z_1)\scalar^{\inn}_{\Delta_2}(z_2)\rest}
    =
    r^{\Delta_{0,12}}
    \xx
    2^{\Delta_{1,02}-1}
    m_1^{\Delta_{2,1}+\delta -2 n+2}
    \mg*{
        \half (\Delta_{1,2}-\delta +2 n-2)
        ,
        \half (2 n -\Delta_{12}-\delta)
    }{
        \Delta_1
    }
    \nn
    \\
    \label{eq: 2toN process - OPE limit - MOM contribution}
    &\peq
    \xx
    \intrange{\frac{d\Delta_{0}}{2\pi i}}{1-i\oo}{1+i\oo}
    f^{\inn}_{\Delta_0,\Delta_{i}}(z_{2},z_{i})
    \mg*{
        \half \Delta_{12,0},
        \half (\Delta_{012}-2),
        \half \Delta_{02,1},
        \half \Delta_{01,2}
    }{
        -\half \delta -\half \Delta_0+n,
        \Delta_0-1,
        \half (\Delta_0-\delta +2 n-2)
    }
    +
    \SmallO(r^{\Delta_{0,12}})
    \, .
\end{align}

\paragraph{OPE analysis.}
We combine the two contributions \eqref{eq: 2toN process - OPE limit - TOM contribution} and \eqref{eq: 2toN process - OPE limit - MOM contribution} to the regular celestial amplitude \eqref{eq: 2toN process - regular celestial amplitude}. To analyze the OPE behavior, we need to enclose the $\Delta_{0}$-contour to the right of the principal series.

The Gamma factors in \eqref{eq: 2toN process - OPE limit - TOM contribution} and \eqref{eq: 2toN process - OPE limit - MOM contribution} provide two types of $\Delta_{0}$-poles to the right: for $k\in\NN$, $\Delta_{0} = \delta -2 n+4 + 2 k$ and $\Delta_{0} = \Delta_{12} + 2 k$. These two series are universally present for any \TwoToN scattering process and are independent of the remaining \OneToN data in $f^{\inn}_{\Delta_{0},\Delta_{i}}$.
The second series of poles have appeared in the previous work \cite{Liu:2024lbs,Liu:2025dhh}, which are left for future investigation.
For the first series of poles, we perform the $\Delta_{0}$-integral and pick up the leading pole $\Delta_{*} = \delta -2 n+4$, and then take the massless limit $m_1\to0$. Counting the factor $\mlimit_{\Delta_{1}}^{m_{1}}$ in \eqref{eq: 2toN process - regular celestial amplitude}, the $m_{1}$-dependence becomes
\begin{equation}
    \lim_{m_{1}\to 0}
    m_1^{\Delta_{12}+\delta -2 n}
    =
    \deltaK*{\Delta_{12}+\delta -2 n}
    \, ,
\end{equation}
which implies $\Delta_{*} = \delta -2 n+4 = 4-\Delta_{12}$.
The corresponding contribution to the regular celestial amplitude \eqref{eq: 2toN process - regular celestial amplitude} reads
\begin{equation}
    \label{eq: TOM leading contribution}
    \vev{\scalar^{\inn,\reg}_{\Delta_1}(z_1)\scalar^{\inn}_{\Delta_2}(z_2)\rest}
    \sim
    r^{\Delta_{*,12}}
    \xx
    \frac{1}{8\pi}
    \mg*{2-\Delta_1, 2-\Delta_2}{3-\Delta_{12}}
    \deltaK*{\delta-\Delta_{*}-2n+4}
    f^{\inn}_{\Delta_{*},\Delta_{i}}(z_{2},z_{i})
    \, ,
\end{equation}
and using the the \OneToN regular celestial amplitude \eqref{eq: 1toN process - regular celestial amplitude}, it can be rewritten as
\begin{equation}
    \vev{\scalar^{\inn,\reg}_{\Delta_1}(z_1)\scalar^{\inn}_{\Delta_2}(z_2)\rest}
    \sim
    r^{\Delta_{*,12}}
    \xx
    \frac{1}{4\pi}
    \mg*{2-\Delta_1, 2-\Delta_2}{3-\Delta_{12}}
    \vev{\wave{\scalar}^{\inn,\reg}_{\Delta_{*}}(z_{2})\rest}
    \, ,
\end{equation}
which implies the shadow exchange in the scalar OPE
\begin{equation}
    \scalar^{\inn}_{\Delta_1}\scalar^{\inn}_{\Delta_2}
    \sim
    \frac{1}{4\pi}
    \mg*{2-\Delta_1, 2-\Delta_2}{3-\Delta_{12}}
    \wave{\scalar}^{\inn}_{4-\Delta_{12}}
    \, .
\end{equation}
%%%%%%%%%%%%%%%%%%%%%%%%%%%%%%%%%%%%%%%%%%%%%%%%%%%
\section{Examples}\label{sec:five_point_check}
%%%%%%%%%%%%%%%%%%%%%%%%%%%%%%%%%%%%%%%%%%%%%%%%%%%%%%%%%%%%%%%%%%%%%%%%%%%%%%%%%%
In the previous sections, we argued from OPE consistency that the ordinary Mellin-basis OPE must be completed by shadow-basis exchanges. We also derived the same shadow exchange from the general structure of tree-level $\TwoToN$ regular celestial amplitudes. In this section we illustrate these general statements in explicit examples.

We first study a concrete $\TwoToThree$ scattering process in massless $\phi^3$ theory. This example provides the simplest nontrivial higher-point setting in which the shadow exchange can be extracted directly from a comb-channel regular celestial amplitude. In particular, we show that the pole responsible for the $12$-channel shadow exchange appears at the expected shadow dimension, and that its residue factorizes into the shadow OPE coefficient times the remaining lower-point regular celestial amplitude.

We then discuss spinning examples. A direct $n$-point derivation for gluons and gravitons would require a spinning split representation, which we leave for future work. Nevertheless, four-point gluon and graviton MHV regular celestial amplitudes already provide useful checks. In these cases, the collinear blocks associated with the two OPE limits can be organized into the Mellin block and its shadow block, in agreement with the shadow-completed OPE analysis above.
%%%%%%%%%%%%%%%%%%%%%%%%%%%%%%%%%%%%%%%%%%%%%%%%%%%%%%%%%%%%%%%%%%%%%%%%%%%%%%%%%
\subsection{\TwoToThree scattering in scalar theory}
%%%%%%%%%%%%%%%%%%%%%%%%%%%%%%%%%%%%%%%%%%%%%%%%%%%%%%%%%%%%%%%%%%%%%%%%%%%%%%%%%
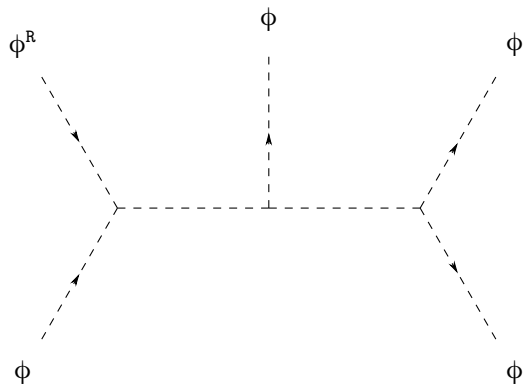
\includeImage{celestial-2-3}{Comb-channel tree diagram for the \TwoToThree scalar example.}

To check the shadow OPE analysis of Section \ref{sec: 2toN process - shadow exchange}, we now consider a concrete \TwoToThree example in massless $\phi^3$ theory. The leading contribution to the amplitude is given by the tree-level exchange diagram in Figure \ref{fig: celestial-2-3}.%
\footnote{
    We have suppressed the $\ii$-factor in the propagator and the coupling constant for simplicity.
}
\begin{equation}
    \cT_{\TwoToThree}
    =
    \frac{1}{(q_1+q_2)^2}
    \frac{1}{(q_4+q_5)^2}
    \deltaMC*{-q_1-q_2+q_3+q_4+q_5}
    \, .
\end{equation}

Following the technique developed in \cite{Liu:2024lbs}, for each of the two propagators we insert a split representation \eqref{eq: split representation}, and due to momentum conservation, only massive parts in \eqref{eq: split representation} survive. Then the mass-regularized celestial amplitudes can be written as an integral over three-point celestial amplitudes, giving
\begin{equation}
    \begin{aligned}
    \vev{\scalar^{\tacp,i m_{1}}_{\Delta_1}\scalar^{\inn}_{\Delta_2}\scalar^{\out}_{\Delta_3}\scalar^{\out}_{\Delta_4}\scalar^{\out}_{\Delta_5}}
    &=
    \intrange{\frac{d\Delta_{\alpha}}{2\pi i}}{1-i\oo}{1+i\oo}
    \intrange{\frac{d\Delta_{\beta}}{2\pi i}}{1-i\oo}{1+i\oo}
    \rho^{\texttt{T}}_{\Delta_{\alpha},\Delta_{\beta}}
    G_{\Delta_{\alpha},\Delta_{\beta}}^{\Delta_{i}}
    \, ,
    \\
    \vev{\scalar^{\inn,m_{1}}_{\Delta_1}\scalar^{\inn}_{\Delta_2}\scalar^{\out}_{\Delta_3}\scalar^{\out}_{\Delta_4}\scalar^{\out}_{\Delta_5}}
    &=
    \intrange{\frac{d\Delta_{\alpha}}{2\pi i}}{1-i\oo}{1+i\oo}
    \intrange{\frac{d\Delta_{\beta}}{2\pi i}}{1-i\oo}{1+i\oo}
    \rho^{\texttt{M}}_{\Delta_{\alpha},\Delta_{\beta}}
    G_{\Delta_{\alpha},\Delta_{\beta}}^{\Delta_{i}}
    \, .
    \end{aligned}
\end{equation}
Here $G$ is the five-point comb-channel conformal block, and the spectral densities are
\begin{equation}
    \begin{aligned}
    \rho^{\texttt{T}}_{\Delta_{\alpha},\Delta_{\beta}}
    &=
    \intrange{dm_{\alpha}}{0}{\oo}
    \intrange{dm_{\beta}}{0}{m_{\alpha}}
    \coThree(
        \scalar^{\tacp,i m_{1}}_{\Delta_{1}}
        \scalar^{\inn}_{\Delta_{2}}
        \scalar^{\out,m_{\alpha}}_{\Delta_{\alpha}}
    )
    I_{\Delta_{\alpha},\Delta_{\beta}}
    \, ,
    \\
    \rho^{\texttt{M}}_{\Delta_{\alpha},\Delta_{\beta}}
    &=
    \intrange{dm_{\alpha}}{m_{1}}{\oo}
    \intrange{dm_{\beta}}{0}{m_{\alpha}}
    \coThree(
        \scalar^{\inn,m_{1}}_{\Delta_{1}}
        \scalar^{\inn}_{\Delta_{2}}
        \scalar^{\out,m_{\alpha}}_{\Delta_{\alpha}}
    )
    I_{\Delta_{\alpha},\Delta_{\beta}}
    \, ,
    \end{aligned}
\end{equation}
where the common factor is
\begin{equation}
    I_{\Delta_{\alpha},\Delta_{\beta}}
    =
    m_{\alpha}^{-2}
    m_{\beta}^{-2}
    \mu^{m_{\alpha}}_{\Delta_{\alpha}}
    \mu^{m_{\beta}}_{\Delta_{\beta}}
    \coShadow_{\DeltaS_{\alpha}}^{\Delta_{3},\Delta_{\beta}}
    \coShadow_{\DeltaS_{\beta}}^{\Delta_{4},\Delta_{5}}
    \coThree(
        \scalar^{\inn,m_{\alpha}}_{\DeltaS_{\alpha}}
        \scalar^{\out}_{\Delta_{3}}
        \scalar^{\out,m_{\beta}}_{\Delta_{\beta}}
    )
    \coThree(
        \scalar^{\inn,m_{\beta}}_{\DeltaS_{\beta}}
        \scalar^{\out}_{\Delta_{4}}
        \scalar^{\out}_{\Delta_{5}}
    )
    \, .
\end{equation}
Then we can perform the $m_{\alpha}$- and $m_{\beta}$-integrals using the same method as Section \ref{sec: 2toN process - shadow exchange}, giving
\begin{equation}
    \begin{aligned}
    &
    \rho^{\texttt{T}}_{\Delta_{\alpha},\Delta_{\beta}}
    =
    2^{\Delta_{1,2345}-4}
    m_1^{\Delta_{2345,1}-6}
    \mg{
        1-\Delta_1,
        \half \Delta_{1\alpha ,2},
        \half \Delta_{2\alpha ,1},
        \half \Delta_{3\alpha ,\beta},
        \half \Delta_{3\beta ,\alpha},
        \half \Delta_{4\beta ,5},
        \half \Delta_{5\beta ,4},
        \half \Delta_{\alpha \beta ,3}
    }{
        \Delta_{\alpha}-1,
        \Delta_{\beta}-1,
        \Delta_{\alpha},
        \Delta_{\beta}
    }
    \\
    &\peq
    \xx
    \mg{
        \half (8-\Delta_{12345}),
        \half (\Delta_{345,\alpha}-4),
        \half (\Delta_{345\alpha}-6),
        \half (\Delta_{3\alpha \beta}-2),
        \half (\Delta_{45,\beta}-2),
        \half (\Delta_{45\beta}-4)
    }{
        \half (4-\Delta_{12\alpha}),
        \half (\Delta_{2345,1}-4),
        \half (\Delta_{345,\alpha}-2),
        \half (\Delta_{345\alpha}-4),
        \half (\Delta_{\alpha ,12}+2)
    }
    \, ,
    \\[1em]
    &
    \rho^{\texttt{M}}_{\Delta_{\alpha},\Delta_{\beta}}
    =
    2^{\Delta_{1,2345}-4}
    m_1^{\Delta_{2345,1}-6}
    \mg{
        \half \Delta_{12,\alpha},
        \half \Delta_{1\alpha ,2},
        \half \Delta_{2\alpha ,1},
        \half \Delta_{3\alpha ,\beta},
        \half \Delta_{3\beta ,\alpha},
        \half \Delta_{4\beta ,5},
        \half \Delta_{5\beta ,4},
        \half \Delta_{\alpha \beta ,3}
    }{
        \Delta_1,
        \Delta_{\alpha}-1,
        \Delta_{\beta}-1,
        \Delta_{\alpha},
        \Delta_{\beta}
    }
    \\
    &\peq
    \xx
    \mg{
        \half (\Delta_{1,2345}+6),
        \half (8-\Delta_{12345}),
        \half (\Delta_{12\alpha}-2),
        \half (\Delta_{3\alpha \beta}-2),
        \half (\Delta_{45,\beta}-2),
        \half (\Delta_{45\beta}-4)
    }{
        \half (\Delta_{345,\alpha}-2),
        \half (8-\Delta_{345\alpha}),
        \half (\Delta_{345\alpha}-4),
        \half (\Delta_{\alpha ,345}+6)
    }
    \, .
    \end{aligned}
\end{equation}
The spectral density entering the regular celestial amplitude is
\begin{equation}
    \rho^{\reg}_{\Delta_{\alpha},\Delta_{\beta}}
    =
    \lim_{m_{1}\to0}
    \thalf
    \mlimit_{\Delta_{1}}^{m_{1}}
    \LR{
        \rho^{\texttt{M}}_{\Delta_{\alpha},\Delta_{\beta}}
        -
        \rho^{\texttt{T}}_{\Delta_{\alpha},\Delta_{\beta}}
    }
    \, .
\end{equation}

Before analyzing the OPE, we also need the integral representation of the remaining \OneToThree celestial amplitude,
\begin{equation}
    \vev{\scalar^{\inn,m_{0}}_{\Delta_0}\scalar^{\out}_{\Delta_3}\scalar^{\out}_{\Delta_4}\scalar^{\out}_{\Delta_5}}
    =
    \intrange{\frac{d\Delta_{\beta}}{2\pi i}}{1-i\oo}{1+i\oo}
    \rho^{\texttt{M}}_{\Delta_{\beta}}
    G_{\Delta_{\beta}}^{\Delta_{i}}
    \, .
\end{equation}
Applying the same method as above, the spectral density is given by
\begin{equation}
    \begin{aligned}
    \rho^{\texttt{M}}_{\Delta_{\beta}}
    &=
    2^{\Delta_{0,345}-3}
    m_0^{\Delta_{345,0}-4}
    \mg{
        \half \Delta_{03,\beta},
        \half \Delta_{0\beta ,3},
        \half \Delta_{3\beta ,0},
        \half \Delta_{4\beta ,5},
        \half \Delta_{5\beta ,4}
    }{
        \Delta_0,
        \Delta_{\beta}-1,
        \Delta_{\beta}
    }
    \\
    &\peq \xx
    \mg{
        \half (\Delta_{03\beta}-2),
        \half (\Delta_{45,\beta}-2),
        \half (\Delta_{45\beta}-4)
    }{
        \half (\Delta_{0345}-4),
        \half (\Delta_{345,0}-2)
    }
    \, .
    \end{aligned}
\end{equation}
For the regular celestial amplitude $\vev{\wave\scalar^{\inn,\reg}_{\Delta_0}\scalar^{\out}_{\Delta_3}\scalar^{\out}_{\Delta_4}\scalar^{\out}_{\Delta_5}}$, since there is no tachyonic contribution due to momentum conservation, the corresponding spectral density is therefore
\begin{equation}
    \rho^{\reg}_{\Delta_{\beta}}
    =
    \lim_{m_{0}\to0}
    \thalf
    \rho^{\texttt{M}}_{\Delta_{\beta}}
    \, .
\end{equation}

We now extract the shadow OPE from the five-point regular celestial amplitude.
When enclosing the $\Delta_{\alpha}$-contour to the right, the pole associated with the shadow OPE comes from the Gamma factor $\gm*{\half(\Delta_{345,\alpha}-4)}$ in $\rho^{\texttt{T}}_{\Delta_{\alpha},\Delta_{\beta}}$ and is located at $\Delta_{*} = \Delta_{345}-4$.
In the massless limit $m_{1}\to0$, the mass dependence provides the overall support
\begin{equation}
    \lim_{m_{1}\to0}
    m_{1}^{\Delta_{12345}-8}
    =
    \deltaK*{\Delta_{12345}-8}
    \, ,
\end{equation}
then the pole becomes the desired conformal dimension of the shadow exchange
\begin{equation}
    \Delta_{*}
    =
    \Delta_{345}-4
    =
    4-\Delta_{12}
    \, .
\end{equation}
Furthermore, the residue of this pole factorizes into the shadow OPE coefficient and the spectral density of the remaining \OneToThree regular celestial amplitude:
\begin{equation}
    \operatorname*{Res}_{\Delta_{\alpha}=\Delta_{*}}
    \rho^{\reg}_{\Delta_{\alpha},\Delta_{\beta}}
    =
    \frac{1}{4\pi}
    \mg*{2-\Delta_1, 2-\Delta_2}{3-\Delta_{12}}
    \eval[\big]{\rho^{\reg}_{\Delta_{\beta}}}{\Delta_{0}=\Delta_{*}}
    \, .
\end{equation}
This reproduces the \TwoToN result of Section \ref{sec: 2toN process - shadow exchange} in the explicit comb-channel expansion.

%%%%%%%%%%%%%%%%%%%%%%%%%%%%%%%%%%%%%%%%%%%%%%%%%%%%%%%%%%%%%%%%%%%%
\subsection{\TwoToTwo scattering in gauge theory and gravity}
%%%%%%%%%%%%%%%%%%%%%%%%%%%%%%%%%%%%%%%%%%%%%%%%%%%%%%%%%%%%%%%%%%%%%

For spinning particles, an analogous derivation requires a spinning split representation for gluons and gravitons. Such a representation should allow one to isolate the spinning shadow exchange directly from $n$-point regular celestial amplitudes and to match the resulting OPE coefficients with those predicted by the shadow-completed collinear OPE. We leave this problem to future work.

Although a general $n$-point spinning example is not yet available, the four-point gluon and graviton MHV regular celestial amplitudes already provide nontrivial checks. These cases were computed in \cite{Liu:2025voe}, where the OPE coefficients were shown to agree with the corresponding collinear OPE.
For a four-point $s$-channel block expansion, the collinear OPE predicts two conformal blocks, associated with the $12$ and $34$ OPE limits. According to the discussion in Section \ref{sec:shadow_states}, the $34$ collinear OPE can be interpreted as the shadow exchange associated with the $12$ channel. This can be checked directly.

For the gluon regular celestial amplitude
\begin{equation}
    \vev{
        \gluon^{\inn,a_1}_{\Delta_1,-}
        \gluon^{\inn,a_2}_{\Delta_2,-}
        \gluon^{\out,a_3}_{\Delta_3,+}
        \gluon^{\out,a_4}_{\Delta_4,+}
    },
\end{equation}
the $34$ OPE exchanges $\gluon^{\out}_{\Delta_{34}-1,+}$ with OPE coefficient $\bt*{\Delta_3-1,\Delta_4-1}$. Using the overall scaling support $\delta(\Delta_{1234}-4)$, one finds that the quantum numbers $(\Delta_{34}-1,+)$ are precisely those of the shadow of the incoming $12$-channel gluon exchange
\begin{equation}
    \shadow[\gluon^{\inn}_{\Delta_{12}-1,-}]
    \, .
\end{equation}
Furthermore, one can verify that the conformal block and coefficient obtained from the $34$ collinear OPE combine with the $12$ collinear block and coefficient into a conformal partial wave, in agreement with the analysis in Section \ref{sec:shadow_states}. In this sense, the four-point MHV amplitude already realizes the same shadow-pairing mechanism at the level of conformal blocks. The corresponding OPE coefficient also matches the coefficient expected for the shadow exchange.

The same check can be carried out for the four-graviton MHV amplitude. These four-point results provide evidence that the shadow-completed OPE is compatible with regular celestial amplitudes beyond the scalar sector.

%%%%%%%%%%%%%%%%%%%%%%%%%%%%%%%%%%%%%%%%%%%%%%%%%%%%%%%%%%%%%%%%%%%%%%%%%
\section{Conclusion and discussion}
%%%%%%%%%%%%%%%%%%%%%%%%%%%%%%%%%%%%%%%%%%%%%%%%%%%%%%%%%%%%%%%%%%%%%%%%%

In this paper, we have investigated the role of shadow-basis operators in the celestial OPE.
The starting point was the observation that a single four-dimensional massless particle admits two natural conformal-primary representatives on the celestial sphere: the usual Mellin-basis operator and its shadow transform. This naturally raises the question of whether the celestial OPE can be consistently formulated using only Mellin-basis operators.

We have argued that the answer is negative. Even if one begins with an operator algebra generated solely by Mellin-basis primaries, consistency of the ordinary OPE forces the inclusion of shadow-basis operators.
The essential point is that the two-point function of ordinary massless celestial primaries in the Mellin basis is contact-supported on the celestial sphere, whereas the OPE limit of a standard celestial correlator may require a non-contact two-point bridge between separated celestial points. This bridge is precisely supplied by the mixed two-point function between a Mellin-basis operator and its shadow.

We first established this mechanism in a three-point function describing the decay of a massive scalar into two massless scalars. We then extended the argument to purely massless correlators by introducing a mass regularization, equivalently by working with regular celestial amplitudes, and further to higher-point scalar correlators by reducing one cluster through successive ordinary collinear OPEs.
These considerations lead to the shadow-completed scalar OPE
\begin{equation}
    \scalar^{\inn}_{\Delta_1}(z_1)
    \scalar^{\inn}_{\Delta_2}(z_2)
    \sim
    \coOPE_{\Delta_{12}-2}\,
    \scalar^{\inn}_{\Delta_{12}-2}(z_2)
    +
    \wave{\coOPE}_{4-\Delta_{12}}\,
    \wave{\scalar}^{\inn}_{4-\Delta_{12}}(z_2)
    \, .
\end{equation}
The shadow term does not correspond to a new bulk particle; rather, it is an alternative celestial representative of the same four-dimensional massless scalar one-particle state. Nevertheless, as a local boundary primary it is not contained in the ordinary Verma module generated by the Mellin-basis state $\scalar_{\Delta}(0)\vac$.
Hence, while the bulk particle content remains unchanged, the local boundary operator algebra must be enlarged from a purely Mellin-basis module to a shadow-completed module.

We have further determined the OPE coefficient of the shadow exchange. This coefficient is not an independent OPE datum; it is fixed by the ordinary collinear OPE coefficient and by the universal shadow factor relating a three-point structure to the shadow transform of one of its legs.
The same OPE-consistency logic applies to gluons and gravitons.

Finally, we have extracted the shadow exchange directly from tree-level $\TwoToN$ scalar regular celestial amplitudes and verified the mechanism explicitly in the five-point regular celestial amplitude. Taken together, this provides an independent check that the shadow term required by OPE consistency is indeed present in regular celestial amplitudes, thereby confirming the consistency of the shadow-completed OPE framework.

Several directions remain open for future investigation.

First, it would be important to extend the regular-celestial-amplitude derivation to $n$-point massless spinning amplitudes. The present work establishes the scalar $\TwoToN$ result by combining mass regularization with the scalar split representation, and it would be natural to generalize this construction to gluons and gravitons.

Second, it would be interesting to test the shadow-exchange mechanism in $n$-point massive celestial amplitudes. One possible approach is to employ a split representation adapted to amplitudes involving massive external particles.
A systematic implementation of this program requires a comprehensive understanding of three-point celestial amplitudes involving massive and tachyonic states \cite{Law:2019glh,Liu:2024lbs,Liu:2025voe}. Such three-point structures enter naturally in the split representation and are essential for extracting the OPE data.
Developing these ingredients would make it possible to investigate whether the shadow-completion mechanism found here for massless OPEs has a direct analogue in the massive celestial setting.

More broadly, the results of this paper suggest that the celestial operator algebra should not be formulated in a purely Mellin-basis language. A single bulk particle gives rise to both Mellin and shadow conformal-primary representatives on the celestial sphere.
The shadow representative is not a new bulk degree of freedom, but it is required for the local boundary OPE and for non-contact factorization in separated OPE limits. Understanding this shadow-completed operator algebra should be an important step toward a more complete formulation of celestial CFT.

%%%%%%%%%%%%%%%%%%%%%%%%%%%%%%%%%%%%%%%%%%%%%%%%%%%%%%%%%%%%%%%%

\section*{Acknowledgements}

WJM is supported by the National Natural Science Foundation of China No. 12405082 and Shanghai Pujiang Program No. 24PJA118.

%%%%%%%%%%%%%%%%%%%%%%%%%%%%%%%%%%%%%%%%%%%%%%%%%%%%%%%%%%%%%%%

\clearpage

\appendix

%%%%%%%%%%%%%%%%%%%%%%%%%%%%%%%%%%%%%%%%%%%%%%%%%%%%%%%%%%%%%%%%

\section{A Conformal integral for OPE}

We follow the convention and methodology of \cite[Appendix C]{Liu:2025voe}.
The one-point integrals are
\begin{align}
    \label{eq: cft one-point integral qp}
    &
    I_{q,p}
    =
    \intt{d^{d}x_0}
    (-\hat{q}_0 \cc p_1 )^{-d}
    =
    2^{1-d} \pi^{\frac{d+1}{2}}
    \frac{1}{\gm{\frac{d+1}{2}}}
    m_1^{-d}
    \, ,
    \\
    \label{eq: cft one-point integral qk}
    &
    I_{q,k}^{\tac{\pm i}}
    =
    \intt{d^{d}x_0}
    (-\hat{q}_0 \cc k_1 )^{-d}_{\tac{\pm i}}
    =
    2^{1-d}
    \pi^{\frac{d+1}{2}}
    \frac{1}{\gm{\frac{d+1}{2}}}
    e^{\mp \half i \pi d}
    m_1^{-d}
    \, .
\end{align}

Now we consider the following conformal integrals
\begin{align}
    \label{eq: conformal integral for OPE - EAdS}
    &
    I_{p}
    =
    \intt{d^d x_{3'}}
    \vevv{\op_{\Delta_{1}}(x_{1})\op_{\Delta_{2}}(x_{2})\op_{\DeltaS_{3'}}(x_{3'})}
    (-\hat{p}_3 \cc \hat{q}_{3'})^{-\Delta_3}
    \, ,
    \\
    \label{eq: conformal integral for OPE - dS}
    &
    I_{k,\tac{s}}
    =
    \intt{d^d x_{3'}}
    \vevv{\op_{\Delta_{1}}(x_{1})\op_{\Delta_{2}}(x_{2})\op_{\DeltaS_{3'}}(x_{3'})}
    (-\hat{k}_3 \cc \hat{q}_{3'})^{-\Delta_3}_{\tac{s}}
    \, .
\end{align}
We summarize the asymptotic behaviors in the limit $r=\abs{x_{1,2}}\to0$ and will provide the derivation later:
\begin{align}
    \label{eq: conformal integral for OPE - EAdS - OPE limit}
    &
    I_{p}
    \sim
    r^{\Delta_{3,12}}
    (-\hat{p}_3 \cc \hat{q}_2)^{-\Delta_3}
    I_{1}
    +
    r^{d-\Delta_{123}}
    (-\hat{p}_3 \cc \hat{q}_2)^{\Delta_3-d}
    I_{2}
    \, ,
    \\
    \label{eq: conformal integral for OPE - dS - OPE limit}
    &
    I_{k,\tacpm}
    \sim
    r^{\Delta_{3,12}}
    (-\hat{k}_3 \cc \hat{q}_2)^{-\Delta_3}_{\tacpm}
    I_{1}
    +
    r^{d-\Delta_{123}}
    \LR{
        (-\hat{k}_3 \cc \hat{q}_2)^{\Delta_3-d}_{\tacpm}
        I_{2,1}
        +
        (-\hat{k}_3 \cc \hat{q}_2)^{\Delta_3-d}_{\tacmp}
        I_{2,2}
    }
    \, ,
\end{align}
where the coefficients are
\begin{equation}
    \begin{aligned}
    &
    I_{1}
    =
    \pi^{d/2}
    \mg*{\half d - \Delta_3, \half \Delta_{13,2}, \half \Delta_{23,1}}{\Delta_3, \half (\Delta_{1,23}+d), \half (\Delta_{2,13}+d)}
    \, ,
    \\
    &
    I_{2}
    =
    \pi^{d/2}
    \mg*{\Delta_3-\half d}{\Delta_3}
    \, ,
    \\
    &
    I_{2,1}
    =
    \pi^{d/2}
    \mg*{1-\Delta_3, \Delta_3-\half d}{1-\half d, \half d}
    \, ,
    \\
    &
    I_{2,2}
    =
    \pi^{d/2}
    \mg*{1-\Delta_3}{\half d-\Delta_3+1}
    \, .
    \end{aligned}
\end{equation}

We use two independent methods to derive the EAdS case \eqref{eq: conformal integral for OPE - EAdS - OPE limit}: direct computation; blow-up method. For the dS case \eqref{eq: conformal integral for OPE - dS - OPE limit}, the closed-form of $I_{k,\tacpm}$ is more involved, and we only derive the asymptotic behavior by the blow-up method.

\subsection{EAdS case}

\paragraph{Direct computation of $I_{p}$.}
For the integral $I_{p}$, we first perform the Feynman parametrization for $\Re(\Delta_i)>0$,
\begin{align}
    \label{eq: Feynman-Schwinger parameterization}
    \prod_{i=1}^{n} A_i^{-\Delta_i}
    &=
    \mg*{
        \sum_{i=1}^{n} \Delta_i
    }{
        \Delta_{1},\dotsc,\Delta_{n}
    }
    \LRb{\prod_{i=2}^n \intrange{d\alpha_{i}}{0}{\oo}\alpha_i^{\Delta_i-1}}
    \LRb{A_1+\sum_{i=2}^n \alpha_i A_i}^{-\sum_{i=1}^{n} \Delta_i}
    \, ,
\end{align}
to combine three denominators, giving
\begin{equation}
    \label{eq: I_p Feynman parametrization}
    \begin{aligned}
    I_{p}
    &=
    2^{\half (\Delta_{12,3}+d)}
    \mg*{d}{\Delta_3, \half (\Delta_{1,23}+d), \half (\Delta_{2,13}+d)}
    (-\hat{q}_{12})^{\half (d-\Delta_{123})}
    \\
    &\peq\xx
    \intrange{ds}{0}{\oo}
    \intrange{dt}{0}{\oo}
    \intt{d^d x_{3'}}
    t^{\Delta_3-1}
    s^{\half (\Delta_{2,13}+d-2)}
    \LR{
        (-t \hat{p}_3-\hat{q}_1-s \hat{q}_2) \cc \hat{q}_{3'}
    }^{-d}
    \, ,
    \end{aligned}
\end{equation}
where $\qhat_{ij} \eqq \qhat_{i} \cc \qhat_{j}$.
Since $-t \hat{p}_3-\hat{q}_1-s \hat{q}_2$ is timelike, the $x_{3'}$-integral takes the form of one-point EAdS-type integral \eqref{eq: cft one-point integral qp}, giving
\begin{equation}
    \label{eq: I_p after x3 integral}
    \begin{aligned}
    I_{p}
    &=
    \pi^{\half(d+1)}
    2^{\half (\Delta_{12,3}-d+2)}
    \mg*{d}{\half(d+1), \Delta_3, \half (\Delta_{1,23}+d), \half (\Delta_{2,13}+d)}
    (-\hat{q}_{12})^{\half (d-\Delta_{123})}
    \\
    &\peq\xx
    \intrange{ds}{0}{\oo}
    \intrange{dt}{0}{\oo}
    s^{\half (\Delta_{2,13}+d-2)}
    t^{\Delta_3-1}
    (-2 s t \hat{p}_3 \cc \hat{q}_2-2 t \hat{p}_3 \cc \hat{q}_1-2 s \hat{q}_{12}+t^2)^{-d/2}
    \, .
    \end{aligned}
\end{equation}
Then the $s$-integral gives a Beta function and the remaining $t$-integral gives a hypergeometric function,
\begin{equation}
    \begin{aligned}
    I_{p}
    &=
    2^{\Delta_2}
    \pi^{d/2}
    \mg*{\half \Delta_{13,2}, \half \Delta_{23,1}}{\half d, \Delta_3}
    (-\hat{q}_{12})^{-\Delta_2}
    (-\hat{p}_3 \cc \hat{q}_1)^{\Delta_{2,1}}
    \\
    &\peq\xx
    \Fba{
        \thalf (\Delta_{2,13}+d),
        \thalf \Delta_{23,1},
        \thalf d,
        1+2 \hat{p}_3 \cc \hat{q}_1 \hat{p}_3 \cc \hat{q}_2 \hat{q}_{12}^{-1}
    }
    \, .
    \end{aligned}
\end{equation}
Finally the asymptotic behavior \eqref{eq: conformal integral for OPE - EAdS - OPE limit} in the OPE limit can be extracted by the transformation of hypergeometric function
\begin{equation}
    \begin{aligned}
    \Fba{a,b,c,z}
    &=
    (1-z)^{-a} \mg*{b-a, c}{b, c-a}
    \Fba{a, c-b, a-b+1, (1-z)^{-1}}
    \\
    &\peq
    -(1-z)^{-b} \mg*{a-b+1, b-a, c}{a, -a+b+1, c-b}
    \Fba{b, c-a, -a+b+1, (1-z)^{-1}}
    \, .
    \end{aligned}
\end{equation}

\paragraph{Blow-up method for $I_{p}$.}
To extract the term $r^{d-\Delta_{123}}$ from $I_{p}$, we take the limit $r\to0$ inside the integral of \eqref{eq: I_p after x3 integral}, then the $s,t$-integrals give Beta functions, giving the coefficient $I_{2}$ in \eqref{eq: conformal integral for OPE - EAdS - OPE limit}.

To read off the other term $r^{\Delta_{3,12}}$, we first perform the change of variables $t\to -t \hat{q}_{12}$ to \eqref{eq: I_p after x3 integral},
\begin{equation}
    \begin{aligned}
    I_{p}
    &=
    \pi^{\half(d+1)}
    2^{\half (\Delta_{12,3}-d+2)}
    \mg*{d}{\half(d+1), \Delta_3, \half (\Delta_{1,23}+d), \half (\Delta_{2,13}+d)}
    (-\hat{q}_{12})^{\half \Delta_{3,12}}
    \\
    &\peq\xx
    \intrange{ds}{0}{\oo}
    \intrange{dt}{0}{\oo}
    s^{\half (\Delta_{2,13}+d-2)}
    t^{\Delta_3-1}
    \LR{
        -2 s t \hat{p}_3 \cc \hat{q}_2
        -2 t \hat{p}_3 \cc \hat{q}_1
        -t^2 \hat{q}_{12}
        +2 s
    }^{-d/2}
    \end{aligned}
    \, ,
\end{equation}
then we can take the limit $r\to0$ inside the integral, and perform the $s,t$-integrals to get the coefficient $I_{1}$ in \eqref{eq: conformal integral for OPE - EAdS - OPE limit}.

\subsection{dS case}

We first consider the complex type $I_{k,\tac{c}}$ for $c=\pm i$, and after obtaining the asymptotic behaviors, we can get the real type $I_{k,\tacpm}$ by linear combinations, see \cite[Appendix B]{Liu:2025voe}.
Using the Feynman parametrization \eqref{eq: Feynman-Schwinger parameterization}, we can write $I_{k,\tac{c}}$ as
\begin{equation}
    \begin{aligned}
    I_{k,\tac{c}}
    &=
    2^{\half (\Delta_{12,3}+d)}
    \mg*{d}{\Delta_3, \half (\Delta_{1,23}+d), \half (\Delta_{2,13}+d)}
    (-\hat{q}_{12})^{\half (d-\Delta_{123})}
    \\
    &\peq\xx
    \intrange{ds}{0}{\oo}
    \intrange{dt}{0}{\oo}
    \intt{d^d x_{3'}}
    t^{\Delta_3-1}
    s^{\half (\Delta_{2,13}+d-2)}
    (-t \hat{k}_3 \cc \hat{q}_{3'}-s \hat{q}_{23'}-\hat{q}_{13'}+c \varepsilon )^{-d}
    \, .
    \end{aligned}
\end{equation}

\paragraph{Coefficient $I_{2,n}$.}
Similar to the EAdS case, we take the limit $r\to0$ inside the integral,
\begin{equation}
    \begin{aligned}
    I_{k,\tac{c}}
    &\sim
    2^{\half (\Delta_{12,3}+d)}
    \mg*{d}{\Delta_3, \half (\Delta_{1,23}+d), \half (\Delta_{2,13}+d)}
    (-\hat{q}_{12})^{\half (d-\Delta_{123})}
    \\
    &\peq\xx
    \intrange{ds}{0}{\oo}
    \intrange{dt}{0}{\oo}
    \intt{d^d x_{3'}}
    t^{\Delta_3-1}
    s^{\half (\Delta_{2,13}+d-2)}
    \LR{
        v \cc \hat{q}_{3'}+c \varepsilon
    }^{-d}
    \, ,
    \end{aligned}
\end{equation}
where the characteristic vector is
\begin{equation}
    v=-\hat{k}_3 t-\hat{q}_2 s-\hat{q}_2
    \, .
\end{equation}

If $\hat{k}_3 \cc \hat{q}_2>0$, then $v$ is timelike, and the $x_{3'}$-integral is of dS type \eqref{eq: cft one-point integral qk};
if $\hat{k}_3 \cc \hat{q}_2<0$, then $v$ is spacelike for $s>0$ and $t>-2 (s+1) \hat{k}_3 \cc \hat{q}_2$, or $v$ is timelike for $s>0$ and $0<t<-2 (s+1) \hat{k}_3 \cc \hat{q}_2$, and the $x_{3'}$-integral is of dS/EAdS type respectively.
In either case, the remaining $s,t$-integrals give Beta functions. The three contributions are
\begin{equation}
    \begin{aligned}
    \frac{I_{k,\tac{c}}}{(-\hat{q}_{12})^{\half (d-\Delta_{123})}}
    &
    \sim
    \pi^{d/2}
    e^{-\half \pi  c d}
    2^{\half (\Delta_{123}-d)}
    \mg*{\Delta_3-\half d}{\Delta_3}
    (-\hat{k}_3 \cc \hat{q}_2)^{-d+\Delta_3}_{\tacm}
    \\
    &\peq+
    \pi^{d/2}
    e^{-\half \pi  c d}
    2^{\half (\Delta_{123}-d)}
    \mg*{1-\half d, \half d}{
        \half
        (d-2 \Delta_3+2)
        , \Delta_3
    }
    (-\hat{k}_3 \cc \hat{q}_2)^{-d+\Delta_3}_{\tacp}
    \\
    &\peq+
    \pi^{d/2}
    2^{\half (\Delta_{123}-d)}
    \mg*{1-\half d, \half d, \Delta_3-\half d}{\Delta_3, d-\Delta_3, \Delta_3-d+1}
    (-\hat{k}_3 \cc \hat{q}_2)^{-d+\Delta_3}_{\tacp}
    \, .
    \end{aligned}
\end{equation}
Combining this result to the real type $I_{k,\tacpm}$ gives the coefficients $I_{2,1}$ and $I_{2,2}$ in \eqref{eq: conformal integral for OPE - dS - OPE limit}.

\paragraph{Coefficient $I_{1}$.}
Similar to the EAdS case, we perform the change of variables $t\to -t \hat{q}_{12}$ and write $I_{k,\tac{c}}$ as
\begin{equation}
    \begin{aligned}
    I_{k,\tac{c}}
    &=
    2^{\half (\Delta_{12,3}+d)}
    \mg*{d}{\Delta_3, \half (\Delta_{1,23}+d), \half (\Delta_{2,13}+d)}
    (-\hat{q}_{12})^{\half \Delta_{3,12}}
    \\
    &\peq\xx
    \intrange{ds}{0}{\oo}
    \intrange{dt}{0}{\oo}
    \intt{d^d x_{3'}}
    t^{\Delta_3-1}
    s^{\half (\Delta_{2,13}+d-2)}
    \LR{
        v \cc \hat{q}_{3'}+c \varepsilon
    }^{-d}
    \, ,
    \end{aligned}
\end{equation}
where the characteristic vector is
\begin{equation}
    v
    =
    -t\hat{k}_3 \sqrt{-\hat{q}_{12}}
    -s\frac{\hat{q}_2}{\sqrt{-\hat{q}_{12}}}
    -\frac{\hat{q}_1}{\sqrt{-\hat{q}_{12}}}
    \, .
\end{equation}
This characteristic vector has a finite norm in the limit $r\to0$,
\begin{equation}
    v^{2}
    \to
    2 s t \hat{k}_3 \cc \hat{q}_2
    +2 t \hat{k}_3 \cc \hat{q}_2
    -2 s
    \, .
\end{equation}

Now we can take the limit $r\to0$ inside the integral.
If $\hat{k}_3 \cc \hat{q}_2<0$, then $v$ is timelike, and the $x_{3'}$-integral is of EAdS type \eqref{eq: cft one-point integral qp};
if $\hat{k}_3 \cc \hat{q}_2>0$, then $v$ is spacelike for $s>0$ and $t>\frac{s}{(s+1) \hat{k}_3 \cc \hat{q}_2}$, or $v$ is timelike for $s>0$ and $0<t<\frac{s}{(s+1) \hat{k}_3 \cc \hat{q}_2}$, and the $x_{3'}$-integral is of dS/EAdS type respectively.
After performing the $s,t$-integrals, we obtain
\begin{equation}
    \begin{aligned}
    \frac{I_{k,\tac{c}}}{(-\hat{q}_{12})^{\half \Delta_{3,12}}}
    &
    \sim
    \pi^{d/2}
    2^{\half \Delta_{12,3}}
    \mg*{\half (d-2 \Delta_3), \half \Delta_{13,2}, \half \Delta_{23,1}}{\Delta_3, \half (\Delta_{1,23}+d), \half (\Delta_{2,13}+d)}
    (-\hat{k}_3 \cc \hat{q}_2)^{-\Delta_3}_{\tacp}
    \\
    &\peq+
    \pi^{d/2} 2^{\half \Delta_{12,3}}
    e^{-\half \pi  c d}
    \mg*{1-\half d, \half d, \half (d-2 \Delta_3), \half \Delta_{13,2}, \half \Delta_{23,1}}{\Delta_3, \Delta_3, \half (\Delta_{1,23}+d), \half (\Delta_{2,13}+d), 1-\Delta_3}
    (-\hat{k}_3 \cc \hat{q}_2)^{-\Delta_3}_{\tacm}
    \\
    &\peq+
    \pi^{d/2}
    2^{\half \Delta_{12,3}}
    \mg*{1-\half d, \half d, \half \Delta_{13,2}, \half \Delta_{23,1}}{\Delta_3, \half (\Delta_{1,23}+d), \half (\Delta_{2,13}+d), \Delta_3-\half d+1}
    (-\hat{k}_3 \cc \hat{q}_2)^{-\Delta_3}_{\tacm}
    \, .
    \end{aligned}
\end{equation}
Combining this result to the real type $I_{k,\tacpm}$ gives the coefficient $I_{1}$ in \eqref{eq: conformal integral for OPE - dS - OPE limit}.

%%%%%%%%%%%%%%%%%%%%%%%%%%%%%%%%%%%%%%%%%%%%%%%%%%%%%%%%%%%%%%%

\section*{}

\clearpage

%%%%%%%%%%%%%%%%%%%%%%%%%%%%%%%%%%%%%%%%%%%%%%%%%%%%%%%%%%%%%%%

\clearpage

\printbibliography

%%%%%%%%%%%%%%%%%%%%%%%%%%%%%%%%%%%%%%%%%%%%%%%%%%%%%%%%%%%%%%%%
%%%%%%%%%%%%%%%%%%%%%%%%%%%%%%%%%%%%%%%%%%%%%%%%%%%%%%%%%%%%%%%

\end{document}